# Light Elements in the Universe

*Sofia Randich\* and Laura Magrini*

*INAF Osservatorio Astrofisico di Arcetri, Firenze, Italy*

Due to their production sites, as well as to how they are processed and destroyed in stars, the light elements are excellent tools to investigate a number of crucial issues in modern astrophysics: from stellar structure and non-standard processes at work in stellar interiors to age dating of stars; from pre-main sequence evolution to the star formation histories of young clusters and associations and to multiple populations in globular clusters; from Big Bang nucleosynthesis to the formation and chemical enrichment history of the Milky Way Galaxy and its populations, just to cite some relevant examples. In this paper, we focus on lithium, beryllium, and boron (LiBeB) and on carbon, nitrogen, and oxygen (CNO). LiBeB are rare elements, with negligible abundances with respect to hydrogen; on the contrary, CNO are among the most abundant elements in the Universe, after H and He. Pioneering observations of light-element surface abundances in stars started almost 70 years ago and huge progress has been achieved since then. Indeed, for different reasons, precise measurements of LiBeB and CNO are difficult, even in our Sun; however, the advent of state-of-the-art ground- and space-based instrumentation has allowed the determination of high-quality abundances in stars of different type, belonging to different Galactic populations, from metal-poor halo stars to young stars in the solar vicinity and from massive stars to cool dwarfs and giants. Noticeably, the recent large spectroscopic surveys performed with multifiber spectrographs have yielded detailed and homogeneous information on the abundances of Li and CNO for statistically significant samples of stars; this has allowed us to obtain new results and insights and, at the same time, raise new questions and challenges. A complete understanding of the light-element patterns and evolution in the Universe has not been still achieved. Perspectives for further progress will open up soon thanks to the new generation instrumentation that is under development and will come online in the coming years.

Keywords: stars, galaxy, stellar populations, abundances, nucleosynthesis, spectroscopy, stars abundances

## 1 INTRODUCTION

The most abundant isotope of lithium, $^7$Li[1], is the heaviest element created during Big Bang nucleosynthesis (BBN) and its primordial abundance can be used to probe the standard model of cosmology (e.g., Wagoner, 1973; Steigman, 2006, and references therein). Also, Li abundance in the interstellar medium (ISM) increases by a factor larger than 10 during the evolution of the Galaxy, as originally pointed out by Rebolo et al. (1988); this suggests one or more sites of fresh Li production.

$^6$Li, $^9$Be, $^{10}$B, and $^{11}$B are instead not produced in significant quantities neither during BBN nor in stars. They are created by the interaction of energetic Galactic cosmic rays (GCRs) with the

---

[1]Unless differently indicated, hereafter we will refer to Li, without indicating the isotope.

interstellar medium (Reeves et al., 1970; Meneguzzi et al., 1971) that can happen in two different channels, namely, a direct process in which protons and α-particles in GCRs collide with CNO nuclei in the ISM or an inverse process where CNO nuclei instead collide with protons and α-particle in the ISM. Observations and in particular the trends between the abundance and metallicity allow discriminating between the two channels. In the first case, Be and B should behave as secondary elements and their abundances in metal-poor stars should show a linear correlation with the metal content ([Fe/H] or [O/H]) and a slope around two; in the second case, Be and B should behave instead as primary elements and the slope should be around one. Given its origin and due to the fact that during the early stages of Galactic evolution (<0.5–1 Gyr) GCRs were generated and transported on a Galactic scale, the Be production site was widespread; to a first approximation, the Be abundance should thus have one value at a given time over the whole Galaxy (Beers et al., 2000; Suzuki and Yoshii, 2001) and it can be used as an ideal cosmic clock to investigate the formation histories of the different Galactic populations (Pasquini et al., 2004; Pasquini et al., 2005; Smiljanic et al., 2009b).

Lithium, beryllium, and boron are also unique because they are fragile elements and they are destroyed in stellar interiors by fusion reactions at progressively higher temperatures of 2.5–3, 3.5, and $5 \times 10^6$ K, respectively. It is worth mentioning the contribution of electron capture on $^7$Be, which is an important channel for $^7$Be destruction but also produces $^7$Li according to physical conditions and how they affect the $^7$Be lifetime (Simonucci et al., 2013). This is relevant also in the context of the Galactic evolution of Li.

Being so fragile, these elements are depleted from stellar atmospheres whenever a mechanism is present that is able to transport material down to the stellar region where the temperature is high enough for Li/Be/B burning to occur. LiBeB are hence excellent tracers of stellar physics and, in particular, of mixing processes, both the standard ones (i.e., convection) and the non standard ones, in stars in different evolutionary phases, from the pre- to the post-main sequence (MS) phases (e.g., Pinsonneault, 1997; Eggleton et al., 2006; Randich, 2006; Busso et al., 2007; Denissenkov et al., 2009; Lattanzio et al., 2015; Charbonnel et al., 2020; and references therein). Also, given the different destruction temperatures, simultaneous measurements of the abundances of the three elements allow us to reconstruct a "tomography" of stellar interiors and to understand how deep the mixing has extended.

After the MS, during the first dredge-up (FDU) event, due to dilution, surface Li abundances are expected to decrease by a factor from 30 to 60, depending on the initial stellar mass and metallicity (Iben, 1967) and Li abundances of stars on the red giant branch should be A(Li) < 1.5 dex[2] (e.g., Brown et al., 1989; Mallik et al., 2003; Gonzalez et al., 2009). Classical models do not predict any further decreasing trend of Li abundance in the subsequent evolutionary phases; on the contrary, the abundance of Li is observed to decrease again after the luminosity bump on the red giant branch (RGB; e.g., Charbonnel et al., 1998; Gratton et al., 2000; Lind et al., 2009b). The further decrease in Li has been ascribed to several non standard processes, including the presence of thermohaline (double diffusive) instability (e.g., Eggleton et al., 2006, Eggleton et al., 2008; Charbonnel and Zahn, 2007; Sengupta and Garaud, 2018), thermohaline and gravitational mixing (e.g., Stancliffe and Glebbeek, 2008; Angelou et al., 2011; Henkel et al., 2018), magnetic buoyancy in a dynamo process (e.g., Busso et al., 2007; Cristallo et al., 2008; Nordhaus et al., 2008; Denissenkov et al., 2009; Cristallo and Vescovi, 2020), and rotation (e.g., Denissenkov and Tout, 2000; Denissenkov and VandenBerg, 2003; Denissenkov and Herwig, 2004).

In addition to several processes that destroy it, lithium can be created via the so-called Cameron-Fowler mechanism (Cameron and Fowler, 1971), in both low-mass and more massive stars which may represent sources of Li enrichment in the Galaxy (see, e.g., Sackmann and Boothroyd, 1999; Charbonnel and Balachandran, 2000; Negueruela et al., 2020).

Finally, it is now well known and expected based on theoretical arguments that Li depletion in stars depends on both age and mass; in particular, Li is since long considered as an excellent and independent age tracer for both young low-mass stars and older solar-type stars, although, as we will discuss in the following, the Li-age relationship is complex and depends on other parameters. Thanks to this, lithium has been extensively used to identify and characterize young stellar populations, to study ages and age dispersion (hence star formation histories) in star-forming regions (e.g., Tagliaferri et al., 1994; Palla et al., 2005; Sacco et al., 2007; Jeffries et al., 2017; and references therein), to derive cluster ages through the lithium depletion boundary method (LDB, Stauffer, 2000) and to age date stars hosting exoplanets.

Carbon, nitrogen, and oxygen are among the most abundant elements in the Universe (Asplund et al., 2009). They are crucial for several astrophysical fields, including, e.g., the formation of planetary systems and astrobiology (e.g., Lodders and Fegley, 2002; Suárez-Andrés et al., 2016), stellar structure and evolution (e.g., Salaris and Cassisi, 2005; Busso et al., 2007; Charbonnel and Zahn, 2007; Lattanzio et al., 2015; Lagarde et al., 2019), stellar nucleosynthesis (e.g., van den Hoek and Groenewegen, 1997; Meynet and Maeder, 2002), and Galactic chemical evolution (e.g., Chiappini et al., 2003; Vincenzo et al., 2016). However, their origins are still debated and the role in their production in stars with different masses, metallicities, and rotational velocities is still not definitively settled. In addition, they have shown a great potential to investigate stellar evolution, since they can probe the internal structure through the mixing processes in the giant phase, in which the material processed in the interior to the stellar photospheres is dredged up, modifying C and N abundances. Last by not least, C and N abundances in evolved stars are now commonly used to estimate the ages of stars, in a complementary way to other methods, such as isochrone fitting, asteroseismology, or gyrochronology (see, e.g., Salaris et al., 2015; Martig et al., 2016; Masseron et al., 2017; Casali et al., 2019; Hasselquist et al., 2019).

For the reasons briefly summarized above, in spite of the observational difficulties, LiBeB and CNO are among the most studied elements of the periodic table. Huge progress has been achieved in the last few years and it is indeed impossible to summarize in this review the state of the art in all fields. We have selected a number of key topics, while we remind to the recent literature for several issues that we will not cover. The latter include the multiple populations of globular clusters (e.g., Gratton et al., 2004, Gratton et al., 2012; Renzini 2008; D'Orazi et al., 2015; Milone et al., 2017; D'Antona et al., 2019; Milone et al., 2020a, Milone et al., 2020b); the abundances of C, N, and O from emission-line spectra of H II regions and planetary nebulae (e.g., Toribio San Cipriano et al., 2016; Toribio San Cipriano et al., 2017; Esteban et al., 2018, Esteban et al., 2019; Esteban et al., 2020; Stanghellini and Haywood, 2018); post-main sequence Li evolution (e.g., Charbonnel et al., 2020; Deepak and Reddy, 2020; Kumar and Reddy, 2020; and references therein), Li-rich giant stars and their nature (e.g., Casey et al., 2016; Casey et al., 2019; Gao et al., 2019; Gonçalves et al., 2020; Jorissen et al., 2020; Martell et al., 2020; Sanna et al., 2020; Wheeler et al., 2020; Yan et al., 2020); boron abundances in massive stars (e.g., Proffitt et al., 2016, and references therein); light elements in the context of exoplanet hosting stars (e.g., Delgado Mena et al., 2012; Delgado Mena et al., 2014); the lithium depletion boundary (see, e.g., Stauffer, 2000; Lodieu, 2020; Martín et al., 2020; an references therein); $^6$Li/$^7$Li isotopic ratio and its implications for Li nucleosynthesis (e.g., González Hernández et al., 2019, and references therein).

---

[2] A(Li) = log N(Li)/N(H)+12

## 2 OBSERVATIONAL CHALLENGES

Due to the low abundances of LiBeB in stellar atmospheres, they are primarily observed in their respective resonance lines. Furthermore, due to its low ionization potential (5.39 eV), a large fraction of Li in stars is ionized. Since the resonance line of Li II is located in the extreme UV and is not observable, one should rely on the resonance line of neutral Li, a doublet at 670.774 and 670.789 nm. The strength of the doublet depends strongly on the star temperature and, of course, abundance. The Li feature is strong in cool stars with high Li content, while it may be extremely weak in warm stars that have suffered large amounts of depletion; to give an idea, the Li line equivalent width may reach 1 Å or so in cool pre-main sequence (PMS) stars and be as weak as just 2 mÅ in our Sun. Also, the Li doublet is separated by 0.4 Å only from a Fe I line, and it is blended with it when the spectral resolution is not high enough or the star is a rapid rotator. Hence, precise measurements of the equivalent width may need high-spectral resolution and high signal-to-noise ratio (S/N), to resolve and measure such tiny lines. We mention in passing that a subordinated Li I line exists at 610.3 nm which is detectable and measurable for high Li abundances only (Gratton and D'Antona, 1989).

Beryllium and boron observations are considerably more difficult, since their resonance lines are located at 313.042 and 313.106 nm, close to the atmospheric cut-off (Be II), and at 249.68 and 249.77 nm, observable only from space (B I). For this reason, Li, with its main resonant doublet falling in the optical, is the most studied among the three elements, while fewer observations have been obtained for Be and B.

The lines described above are used to measure LiBeB abundances in cool stars, both unevolved and evolved ones. In particular, we mention that Li is fully ionized in stars warmer than about 8500 K; hence the Li I lines can be observed and measured only in stars later than the A2–A3 spectral type (e.g., Takeda et al., 2012, and references therein). With a few exceptions, Be II are normally used for stars later than the F spectral type; due to the severe blending and to the fact that the blending lines may become dominant, Be measurements are difficult (if not impossible) in stars cooler than about 4700 K (Smiljanic et al., 2011; Takeda and Tajitsu, 2014). Of the three light elements, B is the only one that can be observed in hot stars. For these stars, other transitions have indeed been employed in the literature; in particular, the 136.2 nm B II and 206.6 nm B III line have allowed the determination of B in B- and A-type stars (see, e.g., Cunha, 2010; Kaufer, 2010).

As for the analysis and abundance determination, issues are present for all the three elements. Li can be determined by using equivalent widths or spectral synthesis; as mentioned, the Li line may be blended with the close-by Fe line, whose contribution hence should be taken into account. In very cool stars (below ~4000 K), the spectral region becomes full of molecular bands and appropriate analysis should be performed, considering pseudoequivalent widths and corresponding curves of growth (e.g., Palla et al., 2007). Also, correct determination of Li abundances would need nonlocal thermodynamical equilibrium (non-LTE) and 3D effects to be taken into account, as these are important in certain regions of the parameter space (see, e.g., Lind et al., 2009a; Harutyunyan et al., 2018; and references therein).

Both Be II and B I features are in even more crowded spectral regions than Li; deriving the abundances requires the use of spectrum synthesis methods, good knowledge of the atomic and molecular line lists, and correct treatment of the UV opacity; these represent key factors even when determining the Be abundance of the Sun (e.g., Balachandran and Bell, 1998). Like lithium, boron is affected by NLTE effects (Kiselman and Carlsson, 1996) which may significantly impact the inferred abundances and results, while NLTE effects are not important (or cancel each other) for Be (Garcia Lopez et al., 1995). We refer to Garcia Lopez et al. (1995), Primas et al. (1997), Boesgaard et al. (2005), and Takeda et al. (2011) for a more detailed discussion of Be and B abundance determination.

C abundances are determined using lines of different origin, namely, permitted lines of C I, forbidden lines [C I], and molecular lines of $C_2$, CH, and CO. Although there are many [C I] lines in the solar spectrum, the ideal lines for abundance measurements should be weak and unblended, with accurate atomic data and transition probabilities, possibly formed in LTE. There are several lines with such characteristics, including lines in the optical range at 477.5, 505.2, and 538.0 nm and several lines in the near-infrared. Among the forbidden lines, the [C I] 872.7 nm line has been successfully used to measure C abundance since it is weak and it forms in LTE conditions. The $C_2$ Swan bands offer numerous weak lines, which are useful for the computation of C abundance. The CH radical contributes to the spectrum through three electronic transitions, which produce three bands in the blue and violet spectral ranges (at ~420 nm, at ~390 nm, and at ~314 nm). The commonly used lines in the optical range are the CH G-bandhead at 431 nm and the $C_2$ bandhead at 563.5 nm. Finally, the CO lines provide the combined abundance of carbon and oxygen, and they can be used to measure C once the oxygen abundance is measured, e.g., from its [O I] lines, or vice versa.

Most N I lines are blended with weak CN lines, whose contribution can be estimated from other stronger CN lines in the spectra. In giant stars, several CN lines, weak and unblended, allow us to extract the N abundance, given an independent determination of the C abundance, e.g., from atomic lines or from $C_2$ molecular bands. The CN lines provide the best way to obtain the N abundance, since the N I lines are too weak and located in the ultraviolet regions. The strongest CN lines are located in the near-UV (388.3 nm). In addition, the NH lines, whose strongest NH bandhead is located in the near-UV, would allow us to obtain a direct measurement of nitrogen (Pasquini et al., 2008), but they produce discrepant values with respect to N from CN (see, e.g., Spite et al., 2005).

The [O I] lines at 557.7 nm (blended with a $C_2$ doublet and thus less useful to compute O abundance), 630.0, and 636.3 nm are crucial to determine stellar oxygen abundances. Recently, also oxygen recombination lines have been used to measure its abundances, such as O I triplet at 777.1 nm, strongly affected by NLTE (Caffau et al., 2008), and the weak line at 615.8 nm (e.g., Bertran de Lis et al., 2015). Also, the OH lines in the near-UV, at ~330 nm, and in the near-infrared H band and the CO lines, more widely distributed in the spectral range, can be used. The latter, as already said, needs a previous measurement of C abundance to be used to infer O abundance.

The use of molecular bands to measure CNO abundances has been experimented also with *ad hoc* photometric filters to measure CH, NH, CN, and OH. Such observations have been used to reveal multiple stellar populations in globular clusters (Piotto et al., 2015) and they are at the basis of the chromosome map (see Milone et al., 2017). In addition, a molecular band of CH and CN in the optical and near-infrared ranges allows us to derive the isotopic composition, in particular, $^{12}C/^{13}C$ (see, e.g., Smith et al., 2002; Tautvaišienė et al., 2015; Drazdauskas et al., 2016). C, N, and O measurements from molecular bands are usually more viable in the spectra of giants, both because their surface abundance is enriched in C and N and because their lower temperature makes the molecular bands dominant. Recent updates in the use of molecular bands are summarized in Barbuy et al. (2018).

# 3 HISTORICAL OVERVIEW

## 3.1 Lithium

The first measurement of Li in the Sun was obtained by Greenstein and Richardson (1951) who already discovered back at that time that our star has a factor of ~100 lower abundance than what measured in the Earth and meteorites (which are representatives of the material from which the Sun has formed), implying depletion from the stellar surface. To our knowledge, the first quantitative measurement of Li in relatively large samples of stars started in the early sixties by Bonsack and collaborators (Bonsack, 1959; Bonsack and Greenstein, 1960); they found that young T Tauri stars are characterized by very high lithium abundances and that *it seems, then, that basically the stars cooler than the Sun destroy lithium and that the efficiency of destruction is highly variable but increases with decreasing temperature.* Since then, some 2000 papers have appeared in the literature reporting observations of Li in all types of stars and environments (from young T Tauri stars to evolved giants; from field stars to open and globular cluster members; from the thin disc of our Galaxy to its halo) and allowing a broad number of issues to be addressed and solved, as well as opening numerous new questions, some of which remain unanswered. The interest in lithium observations has indeed not decreased during the years and the major current spectroscopic surveys are yielding new results, as we will discuss in the following sections. We list below, necessarily in a schematic way and with a personal perspective, the main findings and open questions arisen since the early studies.

### 3.1.1 Li Depletion in Stars, Stellar Physics, and Mixing Mechanisms

According to the standard theory of stellar evolution, which does not consider the effects of processes such as rotation, magnetic fields, chromospheric activity and starspots, diffusion, mass loss, and mass accretion, surface Li depletion starts during the PMS phases when stars are fully convective or the surface convection zone (SCZ) is very extended. The amount of depletion during the PMS should depend on mass, age, and metallicity, as well as on a number of critical parameters in standard models, like, e.g., convection efficiency (see, e.g., Pinsonneault, 1997; Jeffries, 2006). Stars more massive than the Sun are predicted to deplete little (if any) lithium during the PMS, since the radiative core develops (much) earlier than Li burning is complete and the temperature at the base of the convective envelope becomes too low (see Figure 1 in Jeffries, 2006). No further Li depletion is expected during the MS phases for these objects. Stars with masses similar to the Sun are instead expected to deplete Li in the PMS phases, but depletion should stop around 15 Myr, when about only 40% of the initial Li content is retained. Lower-mass stars remain in the PMS phase longer and reach a higher temperature (and density) at the base of their deeper SCZ, so they are predicted to deplete larger amounts of Li in the PMS phases. Finally, fully convective stars (masses below 0.4 $M_\odot$) start depleting Li very early during the PMS evolution and, depending on their mass, deplete it very quickly (the lower the mass is, the slower Li depletion occurs). The SCZs of solar-type stars become shallower during the MS, so no further Li destruction is expected to occur, while later-type stars, whose SCZs remain deep with hot enough bases, continue depleting lithium also on the MS. As mentioned, the standard theory also predicts that Li depletion should depend on metallicity, since at fixed mass, more metal-rich stars have deeper SCZs, due to the increased opacity (e.g., Deliyannis et al., 1990).

These predictions of standard models are supported by observations in very general and statistical terms: young, PMS stars are significantly more Li rich than older stars of a similar type; within the same cluster, Li abundance declines with decreasing effective temperature (i.e., mass); metal-poor old halo stars have very likely undergone much less Li depletion (if any) than their metal-rich counterparts. However, thanks to the overwhelming amount of Li measurements that have been obtained during the last decades, it is now well established that Li is not only a function of age, mass, and metallicity and its behavior is considerably more complex than initially thought.

First, observations of open clusters with ages around 30–40 Myr, whose solar mass stars have just arrived on the Zero Age Main Sequence (ZAMS), indicate that these stars still maintain their initial Li content, suggesting that the Sun and similar stars have depleted all their lithium during the MS life, rather than during the PMS (e.g., Randich et al., 2001; Jeffries et al., 2009). Second, a comparison of star clusters of different ages indeed shows that solar-type stars deplete Li during the MS, at variance with standard model predictions, but confirming the original finding of Zappala (1972). Third, and equally unexpectedly, substantial Li depletion has been observed in warm MS stars with very thin convective envelopes; more specifically, stars in a narrow temperature range around 6500 K show an abrupt decline in Li with respect to both warmer and cooler counterparts (the so-called Li dip, see, e.g., Boesgaard and Tripicco, 1986; Balachandran, 1990). Fourth, otherwise similar stars with the same temperature in the same cluster may show a large dispersion in Li abundances (a factor of 10 or more); this has been seen both among cool members of young and very young clusters, like the 100 Myr old Pleiades (e.g., Soderblom et al., 1993; Randich et al., 2001), and among solar-type stars in much older clusters (Pasquini et al., 1997; Pace et al., 2012). The Li scatter among young cluster members seems to be linked to the stellar rotation, since rapid rotators generally show higher lithium abundances than stars with lower rotational velocities. Finally, the metallicity dependence is not yet observationally well established and contrasting results have been proposed.

All these observational pieces of evidence indicate that Li depletion is driven by other processes, besides convection, and that non standard physics cannot be neglected in stellar models, since it very likely results in enhanced or inhibited Li depletion, depending on the stellar mass and evolutionary stage. In fact, in parallel with the increasing volume of available Li abundances, a broad variety of models of different complexity and including one or more non standard processes have been proposed. These include mass accretion in the PMS phases, enhanced radii due to magnetic fields and/or starspots, mass loss, atomic diffusion, internal gravity waves, rotation, and angular momentum transport induced mixing, star-planet interaction, and, possibly, a combination of them (see, e.g., Swenson and Faulkner, 1992; Charbonnel and Talon, 2008; Charbonnel and Lagarde, 2010; Baraffe and Chabrier, 2010; Somers and Pinsonneault, 2014; Somers, 2016; Amard et al., 2019; Deal et al., 2020; Dumont et al., 2020; and references therein). As a matter of fact, as of today, a consensus on the main mechanisms driving or inhibiting Li depletion on the PMS and MS phases has not yet been reached.

Since, for a fixed mass, Li depletion increases with time in both the PMS and MS phases, Li is in principle an excellent age indicator (see Soderblom et al., 2014); indeed, timescales for the depletion can be derived and Li measurements have been used to age date stars and to identify age dispersions within young clusters (e.g., Sestito and Randich, 2005; Palla et al., 2005, to cite a few early studies). However, given the uncertainties mentioned above, the observed scatter in Li for stars of the same age, and the effect of non standard mixing processes, the use of Li to derive precise individual stellar ages is not straightforward.

### 3.1.2 Primordial Lithium

As mentioned, $^7$Li is the heaviest element produced in significant amounts during BBN. $^7$Li production is a sensitive function of the baryon-to-photon ratio, it can be estimated in the framework of standard BBN (SBBN), and it can be compared with Li measurements in old stars that have formed during the very early phases of the Galaxy evolution. In a seminal paper, Spite and Spite (1982) found a remarkably constant (irrespective of the temperature and metallicity) Li abundance—the then so-called Spite *plateau*—in a small sample of warm halo stars. Under the assumption that these stars had not suffered any depletion and that no Galactic enrichment had yet taken place, they interpreted the *plateau* value, A(Li) = 2.05 ± 0.15, as the primordial Li abundance. Also, given the factor of ~10 difference between that value and Li abundance measured in young T Tauri stars, they suggested that one or more sources had contributed to the enrichment of Li during the Galaxy lifetime. Indeed, a great fraction of subsequent observational and theoretical studies were stimulated by Figures 5 and 6 in Spite and Spite (1982). The existence of a *plateau* was confirmed by several following studies (e.g., Pinsonneault et al., 1992, Pinsonneault et al., 1999; Ryan et al., 1999; Meléndez and Ramírez, 2004; Bonifacio et al., 2007, to cite a few) based on a much larger number of stars and more accurate, modern analysis. While the reported *plateau* value is not identical in the different studies, the highest estimate is not larger than A(Li) = 2.4, with a more typical value around A(Li) = 2.2 (e.g., Spite et al., 2012). During the years, there have also been claims that stars belonging to the *plateau* were actually characterized by both a dispersion and trends with effective temperature and [Fe/H]. In particular, at very low metallicities ([Fe/H] < −2.7), abundances show a large scatter that gets larger towards decreasing metallicity (the meltdown, Sbordone et al., 2010). While the reasons for this "meltdown" were not understood, it suggested that some Spite *plateau* stars may have depleted their initial lithium. The true challenge, however, occurred with the measurement of the cosmic microwave background (CMB) provided by WMAP and Planck satellites, according to which the expected primordial abundance is A(Li) = 2.7, a factor of about three to four higher than the abundances of the Spite *plateau* (see, e.g., Cyburt et al., 2016; Mathews et al., 2020; and references therein). In other words, the stellar Li measurements are inconsistent with the CMB (and deuterium observations), and the discrepancy is larger than 5σ. This is referred to as the cosmological Li problem. The obvious question is whether the disagreement is due to uncertainties in stellar physics, or rather due to new, non standard physics that modifies SBBN. On the cosmological side, different explanations have been proposed (see, e.g.,

Mathews et al., 2020); at the same time, numerous studies have instead invoked non standard mixing mechanisms that would cause Li depletion in halo stars (e.g., Tognelli et al., 2020, and references therein).

### 3.1.3 Galactic Evolution of Lithium

As noted above, a Galactic source is required to account for the increase from the initial value, being the Spite *plateau* or the SBBN value, to that measured in young, metal-rich populations and in the material from which the Sun formed. An established channel is the spallation of atoms in the interstellar medium by energetic cosmic rays (like for $^6$Li, Be, and B); however, spallation processes integrated along the Galactic life can account for only about 10% of the required amount of $^7$Li. Based on theoretical investigations, many additional possible contributors to the Li enrichment have been suggested, such as Asymptotic Giant Branch (AGB) stars, red giants, supernovae, novae, and, recently, active stars (e.g., Cameron and Fowler, 1971; D'Antona and Matteucci, 1991; Travaglio et al., 2001; Romano et al., 2001; Cescutti and Molaro, 2019; Kelly et al., 2020). So far no firm conclusions have been reached on the site(s) of Li production and this is called the Galactic Li problem. Since different sources have different evolutionary time scales, like for the other elements, the empirical determination of the distribution of Li vs. metallicity and rate of Li increase provides a tool to assess the relative contributions of the sites of enrichment. We note that, since Li is destroyed in stars, at each metallicity, a large dispersion in abundance is seen; it is hence common to assume that the upper envelope of the distribution of Li abundances as a function of [Fe/H], which should be representative of the pristine Li in the ISM, traces the abundance evolution.

In order to put tight constraints on models and sources of Li production, statistically robust and homogeneous data sets are needed, well covering a large metallicity range and different populations. Up to a few years ago, not many observational studies were available, mostly based on small or inhomogeneous samples of few hundreds of stars (Lambert and Reddy, 2004; Ramírez et al., 2012; Delgado Mena et al., 2015). As we will discuss in **Section 4.4.1**, things changed with the advent of large spectroscopic programs.

We conclude this section by noting that the three topics discussed above are tightly linked; a complete understanding of mixing mechanisms and Li depletion in stars of different mass and metallicity would allow a final answer to be put on the cosmological Li problem (at least from the stellar side) and, at the same time, the secure determination of the initial value of Li in the early Galaxy, which is critical in models of Galactic evolution.

## 3.2 Beryllium and Boron

Due to the observational difficulties, much fewer studies have been performed on beryllium and even less for boron. Nevertheless, those studies have allowed important insights to be obtained in the different areas.

### 3.2.1 Stellar Physics

Simultaneous measurements of Li, Be, and B allow tighter constraints to be put on the mixing processes at work in stars, since different non standard processes predict different trends of Be vs. Li depletion (e.g., Deliyannis, 2000). Beryllium observations in the Sun were first performed by Greenstein and Tandberg Hanssen (1954), who found that, at variance with lithium, Be was undepleted in our star. Since then, not only were many additional measurements of beryllium in the Sun performed, but also a number of studies aimed to simultaneously measure lithium, beryllium, and, at times, boron in MS stars in the Milky Way (MW) field and star clusters. Crucially, state-of-the-art high-resolution spectrographs with high near-UV efficiency, like UVES on the ESO VLT and HIRES on Keck, have allowed measurements of Be both in bright stars in close-by clusters and also in fainter members of more distant, old clusters, pioneered by Randich et al. (2002).

We refer to Boesgaard and Tripicco (1986), Boesgaard (2005), and Randich (2010) for detailed reviews of the results of observations of Be to trace stellar mixing in solar-type and warmer stars, while we just summarize here the main findings and issues. Specifically, several studies have clearly indicated that stars warmer than about 6000 K, included those in the Li dip, have depleted some amount of Be, a Be dip is seen, as well as a Be vs. Li correlation; this supports rotation induced mixing as the main driver of Li and Be depletion. On the contrary, cooler, solar-type stars in clusters, like the Sun, do not show, within the errors (admittedly not negligible), any evidence of substantial Be depletion, nor any Be vs. Li correlations. As noted by Randich (2010), stars that cover two orders of magnitude in Li abundances and have different ages and metallicities do share a similar Be content (see Figure 3 in that paper). These results suggest a shallower mixing in solar-type stars; i.e., the mixing is deep enough to cause Li but not Be depletion to occur. As for solar analogs in the field, the results generally agree with those for the clusters (e.g., Takeda and Tajitsu, 2009); note however that Takeda et al. (2011), based on a larger sample, detected a very small fraction of stars that had significantly depleted Be. They also claimed that, although the solar twins in their sample roughly share similar Be abundances, there may be a tendency of higher Be for more rapid rotators, while the Sun itself may have depleted some Be. To conclude, stars warmer than about 6000 K deplete some Be and show a Be vs. Li correlation, while cooler stars are normally undepleted.

Boron provides a further probe of mixing since it survives to greater depths than Li and Be. However, as already mentioned, boron observations are extremely challenging, hence limited to relatively small samples of bright stars. In the context of the use of B as a mixing tracer, studies have been performed mainly by Ann Boesgaard and collaborators using STIS and GHRS on board the *Hubble Space Telescope*; see also Primas et al. (1999). Only bright, stars warmer than the Sun in the field were observed. Stars that were depleted in Be were found to be also B deficient and a correlation between B and Be was evidenced (see, e.g., Boesgaard et al., 2005; Boesgaard, 2005; and references therein), although the slope of the correlations is smaller than that of Be vs. Li.

### 3.2.2 Galactic Evolution of Be and B

As mentioned in **Section 1**, both Be and B are produced by cosmic ray spallation. Investigating their evolution in the Galaxy thus allows constraints to be put on the production mechanism as well as on Galactic evolution in general and the intensity spectrum of cosmic rays as a function of time.

The study on the Galactic evolution of Be started about 40 years ago (Molaro and Beckman, 1984), although measurements of Be in stars were obtained before that (as discussed above). However only in the early 90s, large enough samples of stars (including metal-poor ones) were observed, allowing the evolution with metallicity to be traced. Specifically, Boesgaard and King (1993) found a quadratic dependence (a linear correlation with a slope around 2) of Be abundances vs. [Fe/H] for low-metallicity halo stars, apparently confirming the CR direct spallation process as the main production channel; on the contrary, the study by Gilmore et al. (1992) revealed a different trend, a linear correlation with a slope around 1 between A (Be) and [Fe/H]. Thanks to the advent of high-resolution spectrographs on large telescopes, this early result was confirmed by later studies, based on much larger samples of stars extending to lower metallicities and providing strong support to the inverse spallation or "supernova" scenario (see, e.g., Primas, 2010; Boesgaard et al., 2011; Smiljanic, 2014; Prantzos, 2012; and references therein), although the exact slope may change based on the analyzed samples (see, e.g., Smiljanic et al., 2009b). Also, some authors claimed that the slope with metallicity may be higher when looking at the trend of Be vs. oxygen rather than iron, while Boesgaard et al. (2011) even suggested the possible presence of two slopes for different metallicity regimes. Furthermore, a dispersion around the main relationship was seen for metallicities [Fe/H] above ~−1.5, where the halo-to-disc transition occurs (Primas, 2010).

Although based on much fewer data, a similar linear dependence of boron on metallicity (slope ~1) as for beryllium was also found based on the first B observations in metal-poor stars by Duncan et al. (1992). This was confirmed by later studies (see, e.g., Garcia Lopez et al., 1998, and references therein); interestingly, the possibility of two slopes also for this element was proposed by Boesgaard et al. (2004). Finally, the constant B/Be ratio measured for metal-poor stars clearly supported the idea that both elements were created by the same channel (see discussion in Garcia Lopez et al., 1998).

### 3.2.3 Beryllium as a Cosmochronometer

The idea that beryllium could be used as a cosmochronometer or a cosmic clock was successfully tested by Pasquini et al. (2004) who performed the first measurement of Be in turn-off (TO) stars in a globular cluster. Indeed, Be abundances suggested that the cluster NGC 6397 formed 0.2–0.3 Gyr after the onset of star formation in the Galaxy, in excellent agreement with the cluster age obtained through MS fitting. The result and the hypothesis that Be can be used as a clock for the early formation of the Galaxy were further tested confirmed by Pasquini et al. (2007). Based on these positive tests, a number of studies successfully exploited Be as a cosmochronometer, or the equivalent of a timescale, to study the star formation history in the halo and thick disc of our Galaxy (Pasquini et al., 2005; Smiljanic et al., 2009b), and stellar populations in general. These studies supported the idea that Be can indeed be used to separate different populations within the same Galactic component, which would also explain the observed scatter in Be.

## 3.3 CNO

The modern approach to the study of CNO abundances started with Suess and Urey (1956), whose tables of abundances provided fundamental constraints to theories of nucleosynthesis in stars. Cameron (1968), in his pioneering work, measured the abundances of several elements in the solar photosphere, including C, N, and O (see also Cameron, 1973). Lambert (1968) and Lambert (1978) gave a summary of all accessible atomic and molecular signatures of carbon, nitrogen, and oxygen, identifying the best ones to measure abundances. The historical paper of Lambert (1978) gave an overview of the observational methods and challenges for the CNO elemental abundances.

Significant progress has been achieved since the first determination by Lambert (1978), including detailed physics, as NLTE and 3D modeling (Asplund et al., 2004; Asplund, 2005; Caffau et al., 2008; Asplund et al., 2009). However, while the literature determinations of C and N solar abundances are more stable, the solar oxygen abundance is still under debate. Since its first determination by Lambert (1978), who gave abundances 12 + log (O/H) = 8.92, it has been revised several times, producing progressively lower values. Asplund et al. (2009) recommended a solar oxygen abundance 12 + log(O/H) = 8.69 ± 0.05, but the redetermination of Caffau et al. (2008) with 3D solar model atmospheres yielded again a higher solar abundance 12 + log (O/H) = 8.76 ± 0.07.

The interest in C, N, and O abundances has increased with time, thanks to the multiple applications of their study in astrophysics. There are indeed several open issues and questions still debated, which we briefly address in the following.

### 3.3.1 The Role of C and N Photospheric Abundances to Constrain Stellar Physics, Mixing Mechanisms, and Stellar Age Dating

While stellar physics and evolution are well known and constrained, there are still aspects, as the occurrence of non standard mixing processes after the bump luminosity on the red giant branch, which produce changes in the abundances of light elements (see, e.g., Eggleton et al., 2006; Denissenkov et al., 2009; Busso et al., 2007; Busso et al., 2010; Lattanzio et al., 2015; Lagarde et al., 2019; and references therein) and that still need to be fully understood. A statistically significant sample of stars, as the ones collected by the spectroscopic surveys, e.g., Gaia-ESO, APOGEE, GALAH, and LAMOST (see the next section), spanning broad intervals in stellar properties, as mass and metallicity, is indeed necessary to progress in this field and to distinguish among the different transport mechanisms, as, e.g., cool bottom processing (e.g., Boothroyd et al., 1995), deep diffusive mixing (e.g., Denissenkov et al., 1998), rotation (e.g., Charbonnel, 1995; Denissenkov and Herwig, 2004), thermohaline instability (e.g., Eggleton et al., 2006; Charbonnel and Zahn, 2007; Charbonnel and Lagarde, 2010), and magnetic fields (e.g., Busso et al., 2007, Busso et al., 2010; Palmerini and Maiorca, 2010; Cristallo and Vescovi, 2020).

Understanding the surface abundances of C and N in giant stars has also important implications to our ability to use the [C/N] ratio to trace stellar ages (see, e.g., Masseron and Gilmore, 2015; Martig et al., 2016; Ness et al., 2016; Lagarde et al., 2019; Casali et al., 2019; Hasselquist et al., 2019).

### 3.3.2 The Multiple Nucleosynthesis Sites of the CNO Elements and the Role of Stellar Rotation

Although the main sites of production of CNO are known (massive stars for O and a combination of massive and low- and intermediate-mass stars (LIMS) for C and N) (see **Section 4.3** for details and references and Kobayashi et al. (2020) for a recent review), there are still open issues about the relative importance of the two major sites of production and their contribution to the global evolution of carbon and nitrogen (see, e.g., Gustafsson et al., 1999; Henry et al., 2000; Matteucci and Chiappini, 2003; Bensby and Feltzing, 2006; Mattsson, 2010), including the role of massive low-metallicity stars (e.g., Vincenzo et al., 2016). In addition, stellar rotation and mass loss play an important role in setting the final yields of massive stars, in particular for oxygen, carbon, and nitrogen, with notable implications and effects in modeling the chemical evolution of galaxies (see Meynet et al., 2018, for a review).

The evolution of nitrogen, and in particular of the N/O abundance ratio, has been studied in our Galaxy with observations of samples of stars and H II regions (e.g., Christlieb et al., 2004; Israelian et al., 2004; Carigi et al., 2005; Esteban et al., 2005; Spite et al., 2005; Lyubimkov et al., 2013; Esteban and García-Rojas, 2018; Lyubimkov et al., 2019) to put constraints on both Galactic evolution and nucleosynthesis processes. Chiappini et al. (2005) interpreted the origin and evolution of nitrogen in our Galaxy comparing the observed abundances in the Milky Way Galaxy with their two-infall model (originally developed in Chiappini et al., 1997), similarly to Gavilán et al. (2006), who introduced a primary component in the production of N from intermediate-mass stars to explain to observed abundances. They attributed the dispersion of N/O at a given metallicity to the variation of the star formation rates (SFRs) across the Galactic disc, as confirmed by the work of Mollá et al. (2006). Subsequent works (Chiappini et al., 2006; Kobayashi and Nakasato, 2011) showed that yields that take into account rotation are necessary to explain the N/O vs. O/H relation in our Galaxy. More recently, Vincenzo and Kobayashi (2018b) reproduced the observed trends with inhomogeneous enrichment from AGB stars. Despite the numerous works dedicated to these elements, their origin is still debated and not fully clarified (see also Kobayashi et al., 2020). Large spectroscopic surveys, combined with state-of-the-art chemical evolution models and stellar yields, allow us to take a step forward in understanding the origin of these elements (see, e.g., Magrini et al., 2018; Hayes et al., 2018; Romano et al., 2019; Griffith et al., 2019; Franchini et al., 2020).

## 4 THE LIGHT ELEMENTS IN THE ERA OF SPECTROSCOPIC SURVEYS

In the last decade, a variety of large spectroscopic surveys or big observational programs have been started/completed and they are allowing us to get new valuable insights on stellar physics and Galactic archaeology in general. Those surveys collect spectra characterized by different resolving powers and covering different spectral intervals, reach different limiting magnitudes, and focus on a variety of populations. Some of them are public (i.e., the products are periodically released to the community), while others are not. The main characteristics of the surveys of interest here are summarized in **Table 1**. We briefly discuss below a few important aspects.

Most of these recent large spectroscopic surveys, such as GALAH (Bland-Hawthorn et al., 2018), Gaia-ESO (Gilmore et al., 2012; Randich et al., 2013, hereafter GES), and LAMOST (Xiang et al., 2017), are designed to work in the optical spectral range, which is rich in abundance diagnostics, including the Li doublet, atomic and molecular lines of oxygen, carbon, and nitrogen.

The spectroscopic surveys APOGEE and APOGEE-2 (Majewski et al., 2017; Zasowski et al., 2017) collected spectra of the Milky Way stellar populations in the near-infrared. They have measured molecular transitions to derive oxygen, carbon, and nitrogen abundances (García Pérez et al., 2016), now available also in their public release DR16. New techniques working on low-resolution spectra, for instance, the LAMOST spectra at $R = 1800$, have also been developed to measure elemental abundances, including oxygen (Ting et al., 2017), even when no oxygen lines are present (Ting et al., 2018). These new methods are based on a data-driven approach, in which a sample of higher-resolution spectra, for which stellar parameters and abundances have been determined, is used to create *labels* that can be transferred to lower-resolution spectra, allowing us to determine not only parameters and abundances but also ages and masses, even from blended features using the whole information contained in the spectrum.

Three of the above-quoted surveys, LAMOST, GALAH, and GES, include the Li I line in the covered spectral range and crucially yield lithium abundances for large, statistically significant samples of stars belonging to different stellar populations, including star clusters. We mention in particular that the Gaia-ESO Survey has targeted more than 60 open clusters; they well cover the age-metallicity-distance parameter space and include stars of different spectral types, hence providing the largest, homogeneous available data set for lithium in star clusters. We stress here that, as the following discussion will also highlight, open star clusters represent fundamental tools to address many of the topics presented in this review, since they provide homogeneous samples of stars covering different masses and evolutionary stages, spanning large age intervals, and being located in a different part of the Galaxy.

In the following, we will present some of the main results, from published papers. We stress however that full exploitation of those data sets has still to be completed; thus, we will also anticipate a few preliminary figures based on the latest internal release of GES, iDR6, in order to show its potential. We also note that smaller, but important, spectroscopic programs have been recently performed; findings from these projects will also be discussed, when relevant.

We finally mention that invaluable contributions have come from the exquisite astrometry of the *Gaia* space mission and its DR2 and recent EDR3 data releases (Gaia Collaboration et al., 2016; Gaia Collaboration et al., 2018; Gaia Collaboration et al., 2020), allowing the determination of a number of crucial parameters for the study of the light elements, as well as from new measurements of rotational periods for large numbers of clusters from the ground with a variety of projects, and from space with the CoRoT, Kepler/K2, and TESS missions.

**TABLE 1** | Characteristics of the main large spectroscopic surveys.

| Survey | Spectral coverage | R | Covered populations | $N_{stars}$ | Lim mag | Li | CNO |
|---|---|---|---|---|---|---|---|
| GALAH | 470–790 nm[a] | 28,000 | Discs (clusters) | $10^6$ | 12 < V < 14 | ✓ | ✓ |
| APOGEE | 1.51–1.70 μm | 22,500 | Discs/halo (clusters) | $10^5$ | 7 < H < 13.8 | | ✓ |
| APOGEE-2 | 1.51–1.70 μm | 22,500 | Discs/halo (clusters) | $10^5$ | 7 < H < 13.8 | | ✓ |
| GES-Giraffe[b] | 647–679 nm | 17,000 | Clusters | $10^5$ | V < 19 | ✓ | |
| GES-UVES | 480–680 nm | 47,000 | Discs/bulge/clusters | $10^4$ | V < 16 | ✓ | ✓ |
| LAMOST | 370–910 nm | 1,800 | Discs/anticenter/spheroid | $2.5 \times 10^6$ | R < 19 | ✓ | ✓ |

[a]Spectral range divided into four windows.
[b]HR15N setup is used for the observations of open clusters with F-G-K-M stars; the other GES-Giraffe setups are used for field stars or for hot stars in star clusters.

## 4.1 Stellar Physics and Mixing
### 4.1.1 Pre-main Sequence Evolution

The new spectroscopic surveys and, in particular, the Gaia-ESO Survey have observed statistically significant samples of members of many young clusters and star-forming regions, well covering, for the first time, the age range between a few Myr and 100 Myr; this allows us to obtain further insights on the PMS evolution of lithium, while opening new challenges. We refer to Bouvier et al. (2016), Jeffries et al. (2017), Franciosini et al. (2020), and Bouvier (2020) for detailed discussions and presentations of the results, while we summarize below the main findings.

Bouvier et al. (2016) investigated the lithium-rotation connection, which had been shown to hold for the older Pleiades and other clusters (see **Section 3.1.1**), at very young ages. They found that a dispersion in Li and a relationship between Li and rotation is already present among low-mass members (0.5–1.2 $M_\odot$) of the NGC 2264 cluster (~5 Myr), with faster rotators being more Li rich, by about 0.2 dex, than slowly rotating stars. The difference in lithium between slow and fast rotators is much smaller than what is seen in the older Pleiades, but it suggests that non standard physics is already active during the PMS of low-mass stars. Several possible explanations were proposed by Bouvier et al. (2016), among which planet ingestion, the effect of the accretion history, early angular momentum evolution, or magnetic fields and activity that would result in enhanced radii and reduced Li burning, as originally proposed by Somers and Pinsonneault (2014) and Somers and Pinsonneault (2015) (see also Somers et al., 2020, for a very recent update). More specifically, rotation and magnetism have since long been identified as non standard processes that may also affect the structure of young stars and hence Li depletion (Spruit and Weiss 1986; Martin and Claret, 1996; Ventura et al., 1998; Mendes et al., 1999; D'Antona et al., 2000; Feiden and Chaboyer, 2013; Spada et al., 2018) and the new data sets indeed confirm that this is the case and in principle allow constraints to be put on the models.

Recent additional observational pieces of evidence of the action of non standard physics and the effect of magnetic fields in young stars were provided by Messina et al. (2016) and Jeffries et al. (2017). The second paper pointed out in particular that standard PMS evolutionary models do not reproduce simultaneously the color-magnitude diagram (CMD) and the lithium depletion pattern of the young Gamma Velorum cluster: specifically, while the CMD could be well fitted with an age of ~7.5 Myr, the strong Li depletion observed among M-type stars would imply a significantly larger age. Jeffries et al. (2017) were able to reconcile both the Li pattern and the CMD at a common age of ~18–21 Myr, by assuming that the radius of low-mass stars is inflated by ~10%; following a semianalytical approach, they modified stellar models accordingly, considering (almost) fully convective stars with a simple polytropic structure. This resulted in significant older ages than in the standard scenario and in a significant shift of the center of the lithium depleted region towards lower effective temperatures.

Independent evidence of inflated radii has been also found in older clusters like the Pleiades (e.g., Jackson et al., 2016, Jackson et al., 2018; Somers and Stassun, 2017). The quoted authors suggested that radius inflation is likely linked to the magnetic activity and/or starspots and is responsible for the observed dispersion in Li abundances. Note that, as well known, the level of activity is linked to rotation; hence, this would explain the Li-rotation connection. Indeed, Somers and Stassun (2017) demonstrated a triple correlation between rotation rate, radius inflation, and enhanced lithium abundance. The issue is further explored by Franciosini et al. (2020), who extends the analysis to four open clusters with ages between ~20 and 100 Myr observed within GES (Gamma Vel, NGC 2547, NGC 2451 B, and NGC 2516) and investigate the effect of magnetic activity and starspots on PMS evolution models of lithium depletion as a function of age.

### 4.1.2 The MS Phases

A great contribution to the investigation of the evolution of lithium during the MS phases of stars more massive than the Sun has been provided by the WIYN Open Cluster Survey (WOCS, Mathieu, 2000). This program has started about 20 years ago and the results for a number of open clusters have been published during the years, following the evolution of Li from the MS to the tip of the red giant branch. The observed clusters cover a large range in metallicity but are mostly older than 1 Gyr. The main recent results on Li evolution during the MS have been summarized by Twarog et al. (2020) (see also Deliyannis et al., 2019; Anthony-Twarog et al., 2018a, Anthony-Twarog et al., 2018b). Specifically, the existence of the Li dip has been confirmed in many clusters; additionally, the new data suggested that the blue-, warm-, or high-mass side of the Li dip is very sharp (they defined it as the "cliff") and its position (magnitude or temperature) does not appear to change with the cluster age; the mass at which the cliff is seen is determined by the cluster metallicity, with higher masses for more metal-rich clusters. Stars warmer than the dip presumably retain their initial Li content; on the contrary, stars on the red side of the dip seem to sit on a plateau, with no trend of A(Li) with increasing magnitude (decreasing mass). Rotation and rotational evolution (spindown) appear as the critical factors for the appearance of the Li dip and its boundaries.

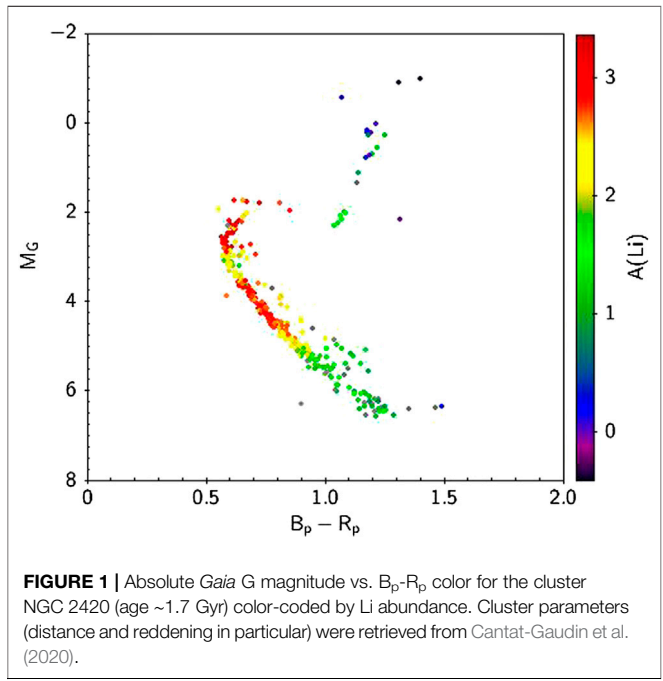

**FIGURE 1** | Absolute *Gaia* G magnitude vs. $B_p$-$R_p$ color for the cluster NGC 2420 (age ~1.7 Gyr) color-coded by Li abundance. Cluster parameters (distance and reddening in particular) were retrieved from Cantat-Gaudin et al. (2020).

As for cooler, lower-mass stars, in intermediate-age clusters, recent WOCS observations of M35 have shown that this cluster behaves similarly to the younger Pleiades, since a dispersion in Li is seen among cool stars, as well as a relationship with rotation and radii (Anthony-Twarog et al., 2018a; Jeffries et al., 2020); this, on the one hand, provides some support for models with inflated radii and lower Li depletion in the faster rotators (more active stars), as discussed above. At the same time, the authors claim that rotational mixing cannot be excluded: given the saturation of magnetic activity, stars would have similar levels of activity, and Li depletion during the PMS may be inhibited for all of them, while additional mixing would subsequently deplete more Li in slow rotators that undergo angular momentum loss (Jeffries et al., 2020). To summarize, the issue is not settled. While new data have confirmed the dispersion among low-mass stars in clusters and have shown that this dispersion sets in early during the PMS, the final reason for it has not yet been definitively confirmed.

As mentioned, GES and in particular its Li data set for open clusters sampling a broad age and metallicity range (from a few Myr to several Gyr, from −0.5 to +0.3 dex in [Fe/H]) offer a great opportunity to further investigate the evolution of Li in MS stars or different temperature. **Figure 1** is a nice illustration of the evolution of lithium in a cluster color-magnitude diagram (CMD). Specifically, the figure shows the *Gaia* $B_p$-$R_p$ vs. G diagram for NGC 2420 (age around 1.7 Gyr, slightly subsolar metallicity), highlighting the changes in Li abundances for stars of different masses along the MS and after it. The dip is clearly visible for stars just below the TO, as well as the decrease of Li along the MS and post-MS evolution.

In **Figures 2–4**, we instead plot Li abundance as a function of absolute G magnitude or effective temperature ($T_{eff}$) for clusters with different ages and/or metallicities. The figures highlight a number of interesting points. First, the dip is clearly visible in all clusters; noticeably, it is already present in the 400 Myr old NGC 3532, hence one of the youngest clusters where the dip is detected. Some scatter is present (possibly due to cluster membership to be refined), but its position, shape, and lowest lithium value in both the A(Li) vs. $T_{eff}$ and A(Li) vs. G diagrams do not greatly depend on metallicity, confirming that the dip occurs at higher masses for more metal-rich clusters (this is just because at a fixed temperature, the mass is higher for more metal-rich stars). On the contrary, the direct comparison of the shape of the dip for NGC 2420 and NGC 3532 (similar metallicities; different ages) may suggest that the dip is shallower at younger ages and its high-mass (blue) side occurs at slightly fainter magnitudes. We also note that presumably undepleted stars on the blue/bright side of the dip seem are slightly more Li rich in the most metal-rich cluster (Ruprecht 134), which has implications for the investigation of the Galactic evolution of lithium (see **Section 4.4.1**).

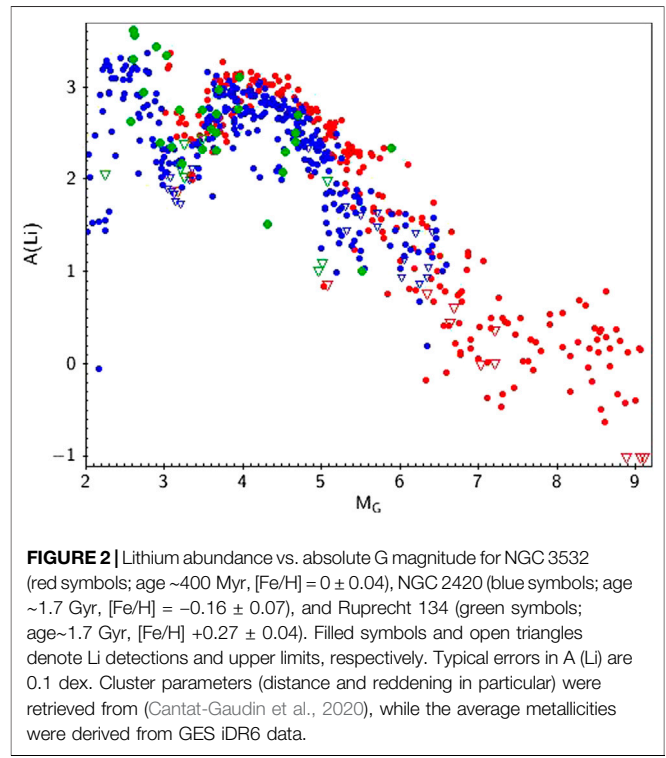

**FIGURE 2** | Lithium abundance vs. absolute G magnitude for NGC 3532 (red symbols; age ~400 Myr, [Fe/H] = 0 ± 0.04), NGC 2420 (blue symbols; age ~1.7 Gyr, [Fe/H] = −0.16 ± 0.07), and Ruprecht 134 (green symbols; age~1.7 Gyr, [Fe/H] +0.27 ± 0.04). Filled symbols and open triangles denote Li detections and upper limits, respectively. Typical errors in A (Li) are 0.1 dex. Cluster parameters (distance and reddening in particular) were retrieved from (Cantat-Gaudin et al., 2020), while the average metallicities were derived from GES iDR6 data.

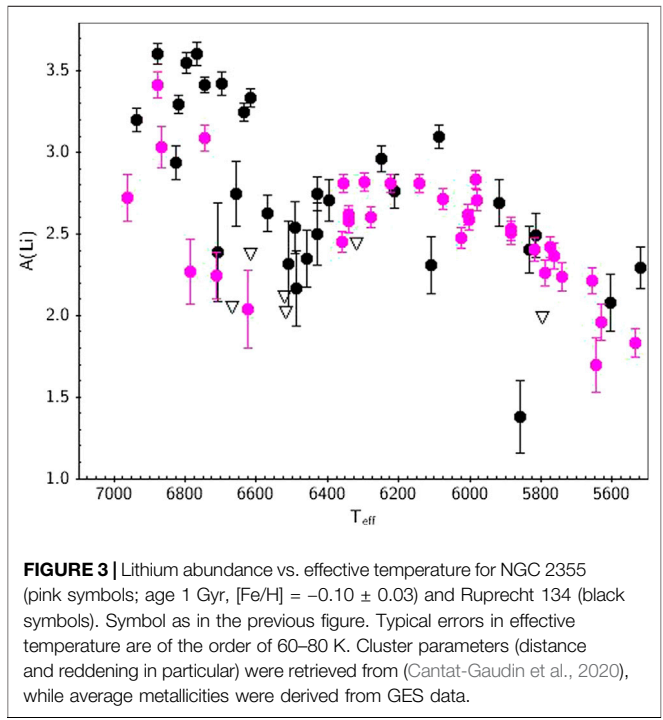

FIGURE 3 | Lithium abundance vs. effective temperature for NGC 2355 (pink symbols; age 1 Gyr, [Fe/H] = −0.10 ± 0.03) and Ruprecht 134 (black symbols). Symbol as in the previous figure. Typical errors in effective temperature are of the order of 60–80 K. Cluster parameters (distance and reddening in particular) were retrieved from (Cantat-Gaudin et al., 2020), while average metallicities were derived from GES data.

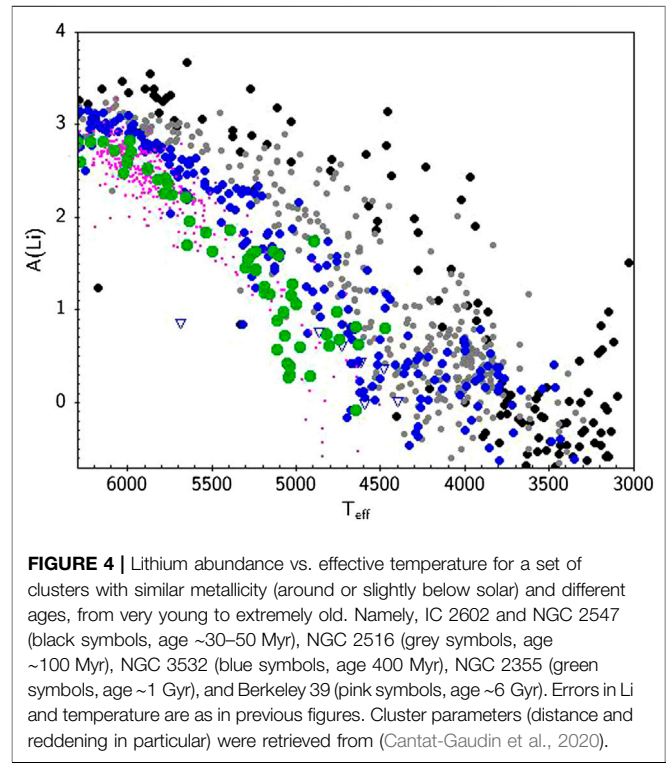

FIGURE 4 | Lithium abundance vs. effective temperature for a set of clusters with similar metallicity (around or slightly below solar) and different ages, from very young to extremely old. Namely, IC 2602 and NGC 2547 (black symbols, age ~30–50 Myr), NGC 2516 (grey symbols, age ~100 Myr), NGC 3532 (blue symbols, age 400 Myr), NGC 2355 (green symbols, age ~1 Gyr), and Berkeley 39 (pink symbols, age ~6 Gyr). Errors in Li and temperature are as in previous figures. Cluster parameters (distance and reddening in particular) were retrieved from (Cantat-Gaudin et al., 2020).

As for stars cooler than the dip, **Figure 4** represents to our knowledge the first homogeneous comparison of the Li vs. temperature distribution from the ZAMS phases up to several Gyrs. The final GES data set includes many more clusters, covering well the age-metallicity parameter space; the figure is therefore anticipation of the excellent potential of this survey to investigate in detail the Li-age evolution, as a function of metallicity and stellar mass (see also **Section 4.2** below). The figure confirms that, for all effective temperatures (or masses), Li depletion is at work on the MS. If we focus on solar-type and warmer stars, for which no depletion on the MS is expected from standard models, the figure indicates a continuous depletion, up to about 1 Gyr, when no more depletion occurs, as the distributions of NGC 2355 (age ~1 Gyr) and Be 39 (age ~6 Gyr) clearly show. Furthermore, all clusters show moderate, if any, dispersion; most importantly, a very minor fraction of cluster members with temperatures similar to the Sun is as Li depleted as it, suggesting that our star has likely had a peculiar Li evolution history. The spread in Li seen among solar-type stars in M67 (discussed earlier in this paper) also seems to be an exception. For lower-mass/cooler stars, the diagram shows the well-known trend of decreasing A(Li) towards lower temperature; depletion increases with age and it becomes faster for lower masses; the scatter within the same cluster appears larger with respect to warmer stars.

GES also provides projected rotational velocities (vsin*i*), which allow further investigation of the Li-rotation relationship. In **Figure 5**, we plot Li abundances vs. the absolute G magnitudes, color-coded by rotational velocity for three clusters with different ages, but similar metallicities. The figure clearly highlights the evolution of the Li dip with time (we note that it is present already at ~100 Myr), the fact that stars in the dip typically have faster rotational velocities than stars outside the dip (in particular, those on the red side of it), and, on average, the Li-rotation connection among low-mass stars (at least in NGC 2516).

As mentioned, the above plots and analysis are preliminary ones, mainly aimed to show the state of the art of Li observations in clusters. A full analysis of the complete data set—we recall that this will also be available to the community through the ESO archive—will provide a detailed view of the Li-age-mass-metallicity-rotation relationships, which will hopefully enable tighter constraints to be put on the models.

### 4.1.3 Beryllium and Boron

Beryllium and boron observations carried out prior to 2010–2015 already exploited available state-of-the-art instrumentation at its best. As a consequence, much fewer studies have been performed in the last five years or so and relatively little progress has been achieved. Among the few recent studies, Desidera et al. (2016) characterized the four Be depleted solar-type stars previously reported by Takeda et al. (2011), finding that all of them are binaries and proposing that the ultra-beryllium depletion may be due to the presence of a brown dwarf companion. Hence, Be depletion seems to be related to very peculiar circumstances, rather than being the normality.

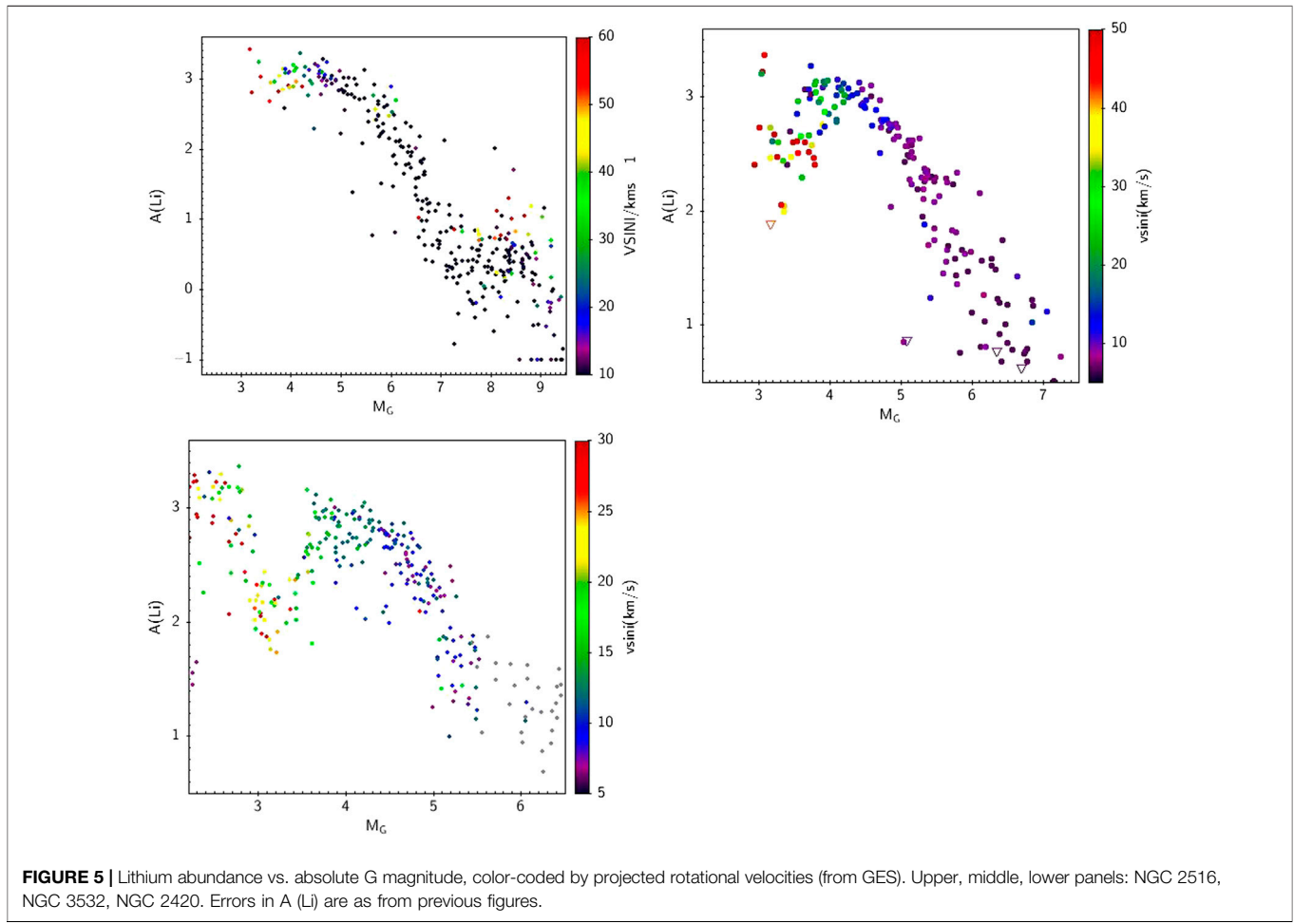

**FIGURE 5** | Lithium abundance vs. absolute G magnitude, color-coded by projected rotational velocities (from GES). Upper, middle, lower panels: NGC 2516, NGC 3532, NGC 2420. Errors in A (Li) are as from previous figures.

As for warmer stars, Boesgaard et al. (2016) determined Li, Be, and B for 79 Hyades dwarfs in the Li dip, confirming both Li and Be dip, as well as detecting a small drop in B abundance across the dip. The B vs. Be correlation is in good agreement with models including rotational mixing. Finally, Boesgaard et al. (2020) very recently determined both Li and Be abundances along the subgiant branch of M67; these stars have evolved from MS stars in the region of the Li and Be dip. They found both a marked decline in Li and a corresponding decrease in Be abundances. They showed that the Be vs. Li pattern could also in this case be well fitted by the models including mixing due to rotation (plus post-MS dilution).

To summarize, a Be vs. Li correlation is now confirmed among F-type stars, probing that, for those stars, the mixing extends down to the Be burning region; the mixing mechanism seems to be shallower and more complex for solar twins, but relatively few data are available, in particular when compared to the wealth of Li measurements from the spectroscopic surveys that we have discussed above.

### 4.1.4 Mixing and Extramixing in Giant Stars

The CNO cycle is at the basis of the production of C, N, and O in stellar interiors. It consists of a set of nuclear reactions, whose final product is the conversion of H in He, and in which C, N, and O atoms play as catalysts (see, e.g., Salaris and Cassisi, 2005). The total amount of C, N, and O is globally unchanged during the CNO cycle. However, the relative abundances of the three elements might evolve with time, driven by the rate of the slowest reaction, which is the one involving a proton capture on $^{14}$N. Thus, at equilibrium conditions, in the stellar core, the abundance of N increases, despite decreasing $^{12}$C abundance and of decreasing in the $^{12}$C/$^{13}$C isotopic ratio. Also, oxygen abundances can be modified, being transformed into N at higher temperatures than C. However, it happens at deeper layers in the stars, which are not reached during the first dredge-up, and consequently, the surface abundance of $^{16}$O is almost unchanged.

At the end of the MS, the amount of CNO-processed material in the core is directly related to the initial stellar mass. More massive stars, reaching higher temperature, would have a larger fraction of the core in which the $^{12}$C can reach burning temperature and thus can be converted in $^{14}$N, with a consequent overabundance in N. After the main sequence, in the ascending phase to the giant branch, during the core contraction, the convective envelope reaches zone in which elements modified by the CNO cycle are present (Iben, 1991), in the so-called first dredge-up (FDU) episode. The FDU is a mixing episode that modifies the surface abundances, since the photospheric abundances are mixed with material enriched in nitrogen and depleted in carbon.

Carbon and nitrogen abundances on the surfaces of evolved stars indeed allow probing their interiors. Their [C/N] ratio, which depends on the mass of the star, can be exploited to derive ages for field red giants (see, e.g., Martig et al., 2016; Ness et al., 2016; Masseron et al., 2017; Casali et al., 2019; and next section) but also to amplitude of the mixing process and the existence of non-canonical processes, called "extramixing," which might be driven by different phenomena, such as rotation, magnetic buoyancy, thermohaline, or gravitational mixing (e.g., Denissenkov et al., 2009; Eggleton et al., 2006; Charbonnel and Lagarde, 2010; Busso et al., 2010; Lattanzio et al., 2015; Cristallo and Vescovi, 2020, among many) and which largely modify also the Li abundance (see **Section 3.3.1**). After the FDU, an extramixing episode occurs at the bump luminosity on the RGB, modifying the photospheric abundances of some elements, including Li, C, and N (e.g., Gilroy, 1989; Gilroy and Brown, 1991; Gratton et al., 2000; Mikolaitis et al., 2012; Smiljanic et al., 2009a; Tautvaišienė et al., 2000; Tautvaišienė et al., 2010; Tautvaišienė et al., 2013; Tautvaišienė et al., 2015; Tautvaišienė et al., 2016). The subsequent dredge-up episodes, namely, the second and third dredge-up (SDU and TDU, respectively), happen in the early-AGB phase, in which the products of the hydrogen burning are brought to the surface at the beginning of the AGB phase (SDU) (see, e.g., Iben and Renzini, 1983; Frost et al., 1998). The AGB phase is characterized by a thermally unstable He-burning shell, with thermal pulses in which the material manufactured in the He-burning shell is convectively mixed and transported close to the base of the H-burning shell. The products of the two shells can be involved in further nucleosynthesis. In addition, after the extinction of the He-burning thermal pulse, the outer convective zone deeply penetrates into the hydrogen and helium burning shells where, eventually, it can transport carbon to the surface (TDU). The TDU might happen several times, possibly producing a C-enhanced star. In what follows, we focus on the comparison between models and observations of RGB stars that have only passed through the FDU.

Using GES data, Lagarde et al. (2019) investigated the effects of thermohaline mixing on C and N observed abundances by performing a comparison with simulations of the observed fields using a model of stellar population synthesis. The C and N abundances in a sample of open and globular clusters, spanning a wide range of ages and metallicities, show the impact of thermohaline mixing at low metallicity. The incidence of thermohaline mixing is indeed able to explain the [C/N] value observed in the lower-mass and older giant stars, as shown their **Figure 6**.

Additional important information on the nature of the mixing processes in evolved stars is related to the carbon isotopic ratio, $^{12}C/^{13}C$. It indeed abruptly decreases when the thermohaline mixing develops during the RGB phase (see, e.g., Charbonnel and Lagarde, 2010). There are several works that investigated the behavior of the C isotopic ratio in star clusters (e.g., Mikolaitis et al., 2012; Drazdauskas et al., 2016; Tautvaišienė et al., 2016), exploiting the advantage of groups of stars that have the same age, metallicity, and origin, from which distance and age and, consequently, their mass and evolutionary status can be precisely determined. The value of $^{12}C/^{13}C$ provides further constraints on the mixing models; in particular, the low values of the carbon isotopic ratio in red clump stars cannot be reproduced without models including extramixing process (see Figure 11 in Lagarde et al., 2019). We recall that thermohaline mixing and all other non standard processes are not derived from first principle physics and they are subject to uncertain parametrization. In this framework, the observations play a crucial role in calibrating and improving their understanding.

## 4.2 Light Elements as Independent Tools to Derive Stellar Ages

The measurements of stellar ages are very challenging, since they cannot be directly obtained from observations (see, e.g., Soderblom et al., 2014; Randich et al., 2018; and references therein). They are usually estimated through the so-called isochrone fitting, i.e., a comparison between observed quantities (magnitudes and colors) or derived quantities (as, e.g., surface gravities, and effective temperatures) and the outcomes of stellar evolution models. The method is more effective for stars belonging to clusters (see, e.g., Randich et al., 2018), for which we can observe several coeval member stars in different evolutionary stages, while it produces large uncertainties for isolated stars, especially if they are located in regions of the plane for which isochrones of different ages are almost overlapping. Other well-established methods are related to asteroseismology, which provide powerful means to probe stellar interiors (see, e.g., Ulrich, 1986; Lebreton and Montalbán, 2009; Soderblom, 2010; Davies and Miglio, 2016; Bellinger, 2019). The oscillation frequencies measured by asteroseismology are indeed closely related to stellar interior properties and tightly linked with the mass and evolutionary state. Comparing the oscillation spectrum with predictions of stellar models, the age and mass of a star can be determined (Lebreton and Montalbán, 2009). However, ages from oscillation spectra are limited by the number of stars observed by dedicated space telescopes (CoRoT, Kepler, PLATO, and TESS). Alternative methods based on the chemical content of stars have been developed in recent years. We discuss here the use of Li and Be abundances and [C/N] ratios to estimate stellar ages.

### 4.2.1 Lithium and Beryllium

As anticipated in **Sections 1** and **3**, Li can in principle be used as an age tracer for both PMS and MS stars. The new spectroscopic programs and Li measurements in several young clusters and associations have provided the possibility to identify members of those clusters based on their high lithium abundance, to measure their age, and to infer the age dispersions (e.g., Murphy and Lawson, 2015; Bravi et al., 2018; Prisinzano et al., 2019; Žerjal et al., 2019, to cite a few). However, as discussed above, PMS depletion is significantly more complex than what standard models predict. The evidence for additional depletion or inhibited depletion for stars of the same age, the effects of non standard processes and other parameters, and the intrinsic difficulty in getting correct Li measurements in PMS stars due to rotation and accretion, eventually are all significant aspects that should be better understood and calibrated and that prevent a precise, quantitative use of Li as an age tracer (e.g., Jeffries, 2017, and references therein).

Similarly, even for stars of similar mass, MS Li depletion is not a simple, continuous function of stellar age only; quantitative use of Li to infer individual stellar ages must hence be done with caution. Inferring ages for field stars similar to the Sun appears particularly important, since it would allow us to age date exoplanet host stars. In this context, a number of recent studies indicate that the so-called solar twins in the MW field deplete lithium by large amounts during their MS life and continue depleting it after the age of 1 Gyr or so (e.g., Meléndez et al., 2014; Carlos et al., 2016; Carlos et al., 2019). Also, in this case, the large number of open clusters observed by GES provides further insights. **Figure 4**, as we have already discussed, suggests that Li depletion in solar-type stars does stop for most of the stars before 1 Gyr, with only a few old cluster members being as depleted as the Sun. In **Figure 6**, we plot the average Li abundance for a number of clusters as a function of age, compared, on a homogeneous basis, to GES field stars. Stars similar to the Sun have been selected following the criteria indicated by Bensby and Lind (2018). The figure clearly indicates that Li depletion among cluster stars increases with increasing age; however, at variance with recent literature results for field stars, but confirming based on more solid statistics the early findings of Sestito and Randich (2005) depletion appears to stop after 1 Gyr. GES MW field stars are all old and, instead, characterized by different amounts of depletion, some of them (actually a minority) being as Li rich as the cluster stars and others being as depleted as the Sun, or even more. While the difference between clusters and field stars is striking and must be further investigated, we conclude that Li is an excellent age indicator for solar twins up to ages around 0.8–1 Gyr; instead, it cannot be easily used as an age tracer for older stars.

We close this section by noting that, after the papers cited in the previous sections aimed to use beryllium as a cosmochronometer, no further studies were performed, due to the limit of current instrumentation to reach fainter targets.

### 4.2.2 Using [C/N] Ratio to Infer Ages of Giant Stars

Several recent works (Masseron and Gilmore, 2015; Salaris et al., 2015; Martig et al., 2016; Ness et al., 2016; Ho et al., 2017; Ho et al., 2017; Lagarde et al., 2017; Feuillet et al., 2018; Casali et al., 2019; Hasselquist et al., 2019) have proposed using the [C/N] abundance ratio measured in the photospheres of evolved stars as an age indicator. In giant stars, the quantity and abundance of the CNO-processed material in the core are proportional to the initial mass. More massive stars, reaching higher temperatures, can convert a larger amount of $^{12}C$ into $^{14}N$. During the episodes of convective mixing, the so-called dredge-ups, the convection reaches its maximum penetration and the processed material is brought to the stellar surface, modifying the surface abundances. The method used to derive stellar ages considers RGB stars that have undergone only the FDU. In RGB stars, the photospheres of massive stars are thus more enriched in N with respect to less massive stars. Since during the RGB phase (short with respect to the MS lifetime), the mass is related to the age, the [C/N] ratio can be used to estimate stellar ages. This is also proven by theoretical stellar models that predict a [C/N] dependence on the mass of the star (hence its age) and, as a second-order effect, on the chemical composition (e.g., Lagarde et al., 2012; Salaris et al., 2015; Lagarde et al., 2017).

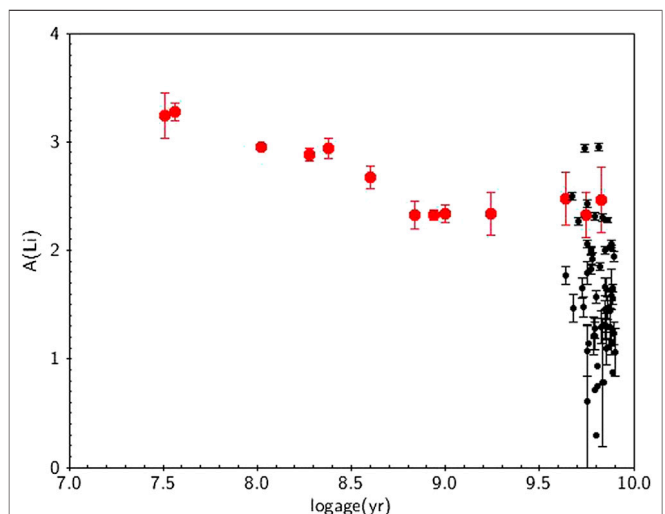

**FIGURE 6** | Average Li abundances as a function of age for solar-type members of a fraction of GES clusters. Thin disc dwarfs in the same temperature range are also shown (black symbols). Solar-type stars are defined as in Bensby and Lind (2018). Error bars for the clusters represent the standard deviation around the average value.

Starting from the work of Masseron and Gilmore (2015), several attempts have been done to provide empirical calibrations of the relationships between stellar ages and the [C/N] abundance ratios. A key point to provide reliable empirical relationships is to have samples of stars with an independent and accurate age determination. A possible source of ages comes from asteroseismic samples of field stars observed by the Kepler and CoRoT satellites (e.g., Anders et al., 2017, for the CoroGEE sample of giant stars). Another possibility is to use the ages of stars clusters, which offer the unique opportunity of well-determined ages through isochrone fitting of many members observed across the cluster sequence (see Casali et al., 2019). In addition, star clusters cover large ranges of ages, distances, and metallicities, and they have been extensively observed by two of the major large spectroscopic surveys, GES and APOGEE. Starting from the ages determined with isochrone fitting for the open clusters observed by Gaia-ESO DR5 and by APOGEE DR14, Casali et al. (2019) calibrated a relationship between cluster age and the [C/N] ratio in their evolved stars, accurately selecting them among post-FDU stars and studying the occurrence of noncanonical mixing. In **Figure 7**, we extend the sample of Casali et al. (2019) including more clusters from the last GES data release. In the figure, we plot [C/N] vs. log(Age (yr)) in star clusters from Casali et al. (2019), which includes both GES DR5 and APOGEE DR14 and in new clusters from GES DR6. The symbols are color-coded by metallicity, indicating a mild dependence on [Fe/H].

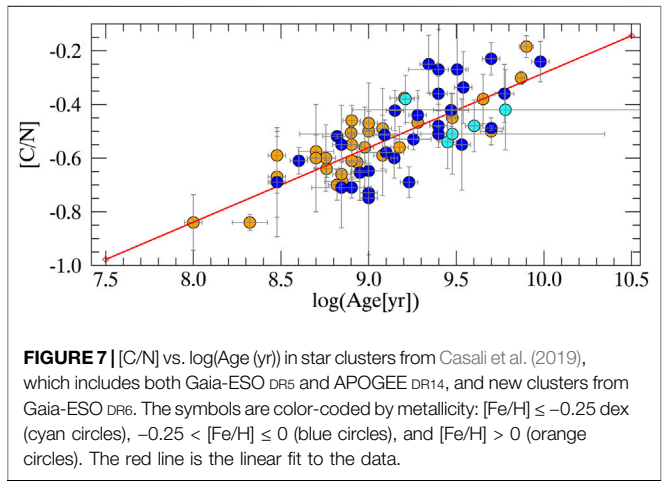

**FIGURE 7** | [C/N] vs. log(Age (yr)) in star clusters from Casali et al. (2019), which includes both Gaia-ESO DR5 and APOGEE DR14, and new clusters from Gaia-ESO DR6. The symbols are color-coded by metallicity: [Fe/H] ≤ −0.25 dex (cyan circles), −0.25 < [Fe/H] ≤ 0 (blue circles), and [Fe/H] > 0 (orange circles). The red line is the linear fit to the data.

As in Figure 8 of Casali et al. (2019), there is a clear relation between the [C/N] abundance ratio and the age of the stars, expressed in logarithmic form. This relationship can be used to infer the ages of giant stars for which we know the evolutionary status and we can measure C and N abundances. The typical uncertainties are in log (age) of ∼0.15–0.20 dex. Although ages from [C/N] can be used only with a statistical meaning, they are excellent alternatives to ages from isochrones. They revealed to be useful to trace the differences in ages between the thin and thick disc populations, as shown, for instance, in Figure 13 of Martig et al. (2016) and in Figure 13 of Casali et al. (2019). The latter is reproduced here in **Figure 8**, where [α/Fe] is plotted as a function of [Fe/H] for field stars in the APOGEE DR14 and Gaia-ESO DR5 samples are shown. At a given [Fe/H], stars belonging to the thick disk have higher [α/Fe], and they are older than thin disc stars as confirmed from the age measured with the [C/N] ratio.

Using the results of the APOGEE survey, Hasselquist et al. (2019) exploited the [C/N] abundance ratio as an age indicator, studying the age-metallicity-abundance trends across the MW disk. Dividing their star sample in age bins, they were able to measure the radial metallicity gradient for the youngest stars (age < 2.5 Gyr) finding a slope equal to −0.060 dex kpc$^{−1}$ in the radial range from 6 to 12 kpc, which is in agreement with other tracers, as H II regions, open clusters, Cepheids, and the CoRoGEE sample (see Anders et al., 2017, and references therein). On the other hand, they found that older stars have a flatter gradient (−0.016 dex kpc$^{−1}$), as predicted by simulations which include stellar migration.

## 4.3 Nucleosynthesis
### 4.3.1 Primordial Nucleosynthesis: The Cosmological Li Problem

The new observations and data collected by the recent spectroscopic observations have shed new light on the cosmological lithium problem but have not solved it. Specifically, Guiglion et al. (2016) in the context of the AMBRE project (Worley et al., 2012; de Laverny et al., 2013) measured Li in a sample of 44 metal-poor stars; they found that these stars share a rather constant lithium abundance (A(Li) = 2.08) with a typical dispersion of 0.22 dex and confirmed that the mean lithium abundance of metal-poor stars in the Galaxy is lower by 0.4–0.5 dex than the SBBN value. For the most metal-poor stars in their sample, they found further evidence for the meltdown.

Very recently, Gao et al. (2020) reported Li abundances for a very large sample (>10$^5$) F-G-K dwarf and subgiant stars observed in the context of GALAH, K2-HERMES, and TESS-HERMES. They divided the sample into "warm" (warmer than the Li dip, see **Section 4.1.2**) and cool stars. The metal-poor ([Fe/H] < −1) cool group very well resembles the Spite plateau, with "low" Li abundances (average around A(Li) = 2.35–2.4); the warm group contains only more metal-rich stars (−1<[Fe/H] <−0.5), which also show a plateau, but with a higher average abundance (A(Li) = 2.69 ± 0.06), roughly consistent with SBBN predictions. The authors of the paper hence claim that these stars have preserved their initial, primordial Li and that both Li depletion and Galactic enrichment have not been significant at these metallicities, thus apparently solving the inconsistency.

The Gaia-ESO Survey has mainly focused on observations of stars in the Galactic discs, with fewer targets in the halo. Also, most of the Milky Way fields have been observed with instrument settings that do not cover the Li line. Nevertheless, GES does eventually include a few halo stars with Li measurements and thus offers an opportunity to further investigate the primordial Li issue, although with a relatively small sample.

In **Figure 9**, we plot Li abundances as a function of metallicity for warm dwarf/TO MW stars ($T_{eff}$ > 6000 K; log > 3.8). The figure shows the usual dispersion for metallicities above [Fe/H] ∼−0.8, due to variable amounts of depletion. Stars with upper limits are typically cooler than stars with Li detections and, as expected, most Li-rich stars in the "metal-rich" regime are very warm and warmer than the dip. On the other hand, with exception of one Li-depleted target, metal-poor stars clearly identify the plateau; GES data do not evidence a cool and a warm (with higher Li) plateau, but, noticeably, one of the plateau stars is relatively warm (warmer than the dip) and shares the same Li abundance as the other stars. We mention in passing that, by considering the nine stars lying on the metal-poor Li plateau, we obtain an average value A(Li) = 2.24 ± 0.12, perfectly consistent with previous values in the literature and still a factor of ∼4 below the cosmological abundance.

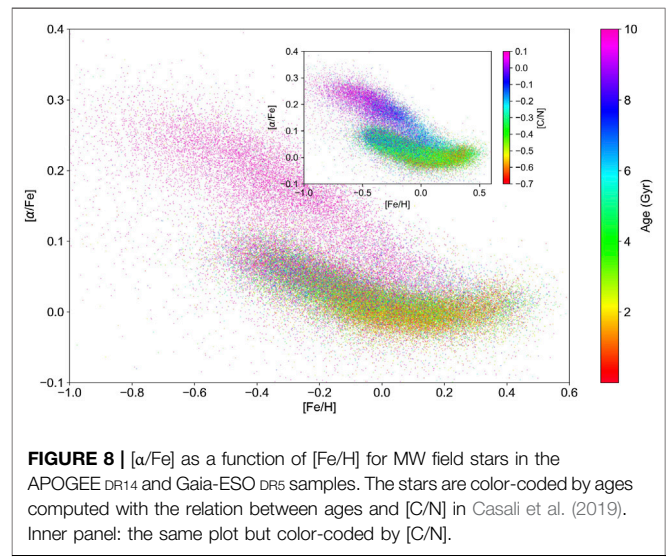

**FIGURE 8** | [α/Fe] as a function of [Fe/H] for MW field stars in the APOGEE DR14 and Gaia-ESO DR5 samples. The stars are color-coded by ages computed with the relation between ages and [C/N] in Casali et al. (2019). Inner panel: the same plot but color-coded by [C/N].

### 4.3.2 The Origin of C, N, and O and the Role of Rotation

Carbon, nitrogen, and oxygen are among the most abundant elements in the Universe. Their abundance ratios give important insights into both stellar physics and Galactic chemical evolution. Their origin, together with other elements, has been extensively discussed in recent reviews and books (see, e.g., Rauscher and Patkós, 2011; Matteucci, 2012; Nomoto et al., 2013; Kobayashi et al., 2020). An effective illustration of the differences in the production sites and time scales of the CNO elements and Fe is in Figure 1 of the review by Maiolino and Mannucci (2019). On one hand, oxygen, like most of the α elements, is produced by massive stars (M > 8 $M_\odot$), which terminate their lives as core-collapse SNe, with short time scales for its release in the interstellar medium. On the other hand, a lower fraction of carbon and nitrogen is produced by massive stars. Most of their production is due to intermediate-mass stars (2 $M_\odot$ < M < 8 $M_\odot$), with longer time scales and thus contributing to the ISM enrichment at later times. In this general overview, there are still several open issues concerning the nucleosynthesis of the CNO elements. As said above, carbon is produced by both massive and LIMS stars. However, the relative importance of the two major sites of production and their contribution to the global evolution of carbon over time is still unclear. For instance, some works (e.g., Gustafsson et al., 1999; Henry et al., 2000) are in favor of a dominant contribution from metal-rich massive stars, while others (e.g., Matteucci and Chiappini, 2003; Bensby and Feltzing, 2006; Mattsson, 2010) find more evidence for the production of C in LIMS during the latest stages of their evolution.

Nitrogen is mostly produced by LIMS, in which it can have both a primary and secondary origin (e.g., van den Hoek and Groenewegen, 1997; Henry et al., 2000; Meynet and Maeder, 2002). The former is produced during the thermal pulses, when some helium products might be transported into the hydrogen burning shell to produce primary nitrogen (Meynet and Maeder, 2002); the latter is the product of the CNO cycle, in which N is formed at expenses of the C and O already present in the star, and thus it increases with metallicity (e.g., Vincenzo et al., 2016). However, to explain the observed plateau in N/O at low metallicity, a production of N in short-living rapidly rotating massive stars is necessary (e.g., Henry et al., 2000; Roy et al., 2020).

Oxygen is almost entirely produced by massive stars and ejected into the interstellar medium via the explosion of core-collapse (type II) SNe. However, convection, rotation, and mass loss play a major role largely affecting the resulting yields (for a review, see Meynet et al., 2018). For instance, the rotation might increase the total metallic yields and in particular the yields of carbon and oxygen by a factor of 1.5–2.5 and of nitrogen over 2-order of magnitude (cf. Hirschi et al., 2004; Chieffi and Limongi, 2013; Limongi and Chieffi, 2018; Meynet et al., 2018).

The effects of rotation on the final yields are even larger at low metallicity and thus particularly important for the nucleosynthesis of the first generations of massive stars in the Universe (see Limongi and Chieffi, 2018, for a recent set of yields with stellar rotation). Rotation appears to drive the internal mixing and trigger several instabilities like shear instabilities or meridional currents. These instabilities are even more efficient at low metallicity, where stars are more compact and the gradients of the angular velocity are, in general, steeper (see Maeder and Meynet, 2001). The changes in the yields affect the results on the global chemical evolution in the Galaxy. Adopting yields from rotating models and making some assumption on the distribution of rotational velocities in massive stars, chemical evolution models for the halo of our Galaxy are able to explain better the observed trends of N/O and C/O (see, e.g., Chiappini et al., 2003; Chiappini et al., 2005; Prantzos, 2019). In particular, the recent work of Prantzos (2019) probed, for the first time, the effect of rotation in the production of N by using a metallicity-dependent distribution of the rotational velocities, rather than a single velocity, and thus reconciling the results of the model with the observations (see their Figure 14.3).

The advent of large spectroscopic surveys made it possible the investigation, based on sizeable statistical samples, of the behavior of carbon in different populations of our Galaxy, and its origin, for which the relative importance of massive and LIMS is still a matter of debate (e.g., Gustafsson et al., 1999; Henry et al., 2000; Chiappini et al., 2003; Matteucci and Chiappini, 2003; Bensby and Feltzing, 2006; Gavilán et al., 2006; Mattsson, 2010; Amarsi et al., 2019; Romano et al., 2019). Recently, the determination of carbon abundance from stellar spectra has been extended to nearby galaxies (see Conroy et al., 2014).

Recent constraints of the origin of C from chemical evolution are presented Franchini et al. (2020), who, thanks to a sample of F-G-K stars from the GES DR5, complemented with *Gaia* DR2 astrometry investigated the behaviors of [C/H], [C/Fe], and [C/Mg] vs. [Fe/H], [Mg/H], and age. The detected trends pose important constraints to the origin of C, suggesting that it is primarily produced in massive stars. In addition, the increase of [C/Mg] for young thin-disk stars might indicate a contribution from low-mass stars, polluting the ISM at later epochs, or increased production from massive stars at high metallicity, due to the enhanced mass loss (see, e.g., Magrini et al., 2017; Prantzos et al., 2020; Chieffi and Limongi, 2020). The GALAH survey is providing detailed abundances for a large sample of stars (70,000 stars in their DR2, see Buder et al., 2018). Griffith et al. (2019) presented C abundances from about 12,000 stars in GALAH DR2, separating the two disc components in the plane [C/Mg] vs. [Mg/H] and suggested that about 75% of solar C comes from core-collapse supernovae, while the remaining fraction is due to other sources with a delayed release of C. Romano et al. (2020) compared the data from both GALAH and GES with the results of Galactic chemical evolution models, concluding that more than 60% of the solar C abundance comes from massive stars. Their results are more generally extended, as they claim that the majority of C in the Universe comes from massive (fast) rotators, with a non-negligible contribution from intermediate-mass stars.

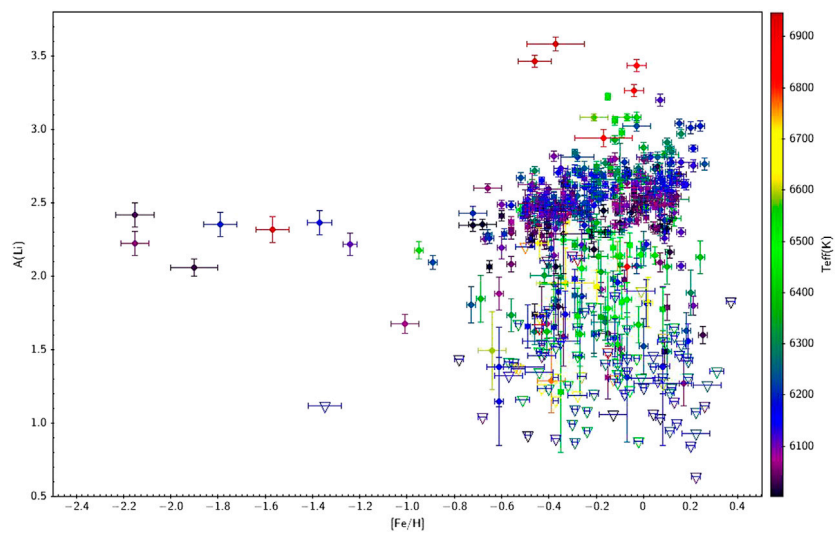

**FIGURE 9** | Li abundances as a function of [Fe/H] for GES MW field stars. Only TO stars are shown. The stars are color-coded by effective temperature. Errors in A(Li) are as in previous figures, while typical errors in [Fe/H] are of the order of 0.05–0.1 dex.

The data from GES are also contributing to our understanding of the evolution of oxygen. Franchini et al. (2021) analyzed the oxygen abundances of a stellar sample representative of the thin and thick discs, aiming at investigating possible differences in their oxygen content and at understanding the origin of the Galactic oxygen enrichment. They found a systematic difference between the [O/Fe] content in the thin and thick disc populations at a given [Fe/H], with the thick disc being more enhanced in [O/Fe] and [O/H] with respect to thin disc stars, and a monotonic decrease of [O/Fe] with increasing metallicity, even at very high metallicity. Their result suggests that the oxygen enrichment is mostly due to massive stars, through core-collapse supernovae, with no evidence of contributions from SNIa or AGB stars. In addition, they found that oxygen and magnesium do not follow the same evolution (see also Magrini et al., 2017), since they do not evolve in lockstep and thus they might have a different origin. Interestingly, they found that the abundance ratio [Mg/O] correlates with stellar ages, and thus it might be used as a good indicator of ages.

Finally, new NLTE abundances for carbon and oxygen have been provided for the GALAH survey (Amarsi et al., 2020). For C and O, the NLTE corrections have minor effects with respect to the LTE abundances. The observed trend in the GALAH survey of [C/Fe] vs. [Fe/H] indicates a decrease that reflects the different timescales for the production of iron (from SNIa) and of carbon (from AGB and massive stars). For O, there is a step linear decrease in [O/Fe] that reflects the two main mechanisms of production of these elements: SNIa for iron and massive stars from oxygen. The results from GALAH are in general good agreement with previous high-spectral resolution observations (Bensby et al., 2014; Amarsi et al., 2019), but they are not completely consistent with the results of the APOGEE survey (Hayes et al., 2018) that indicates a plateau in the [O/Fe] and even a slight increase in [C/Fe] at super solar metallicities. This can be due to the determination of C and O abundances in APOGEE from molecular lines, which can be affected by important systematics (see also Weinberg et al., 2019, for oxygen).

## 4.4 Galactic Evolution
### 4.4.1 Lithium, Be, and B

As mentioned in the introduction, many sources of Li production have been proposed to contribute to the Li enrichment in the Galaxy and to explain the increase from the Big Bang value to the meteoritic value, among which are asymptotic AGB stars ejecta, red giants, supernovae, cosmic ray spallation, and novae. While it is not yet clear which of them provides the dominant contribution, a few recent observational highlights have provided new constraints. In particular, the lines $^7$Li or $^7$Be (which then decay into $^7$Li) have been detected in novae at the early stages, showing that these systems may represent an important source of Li enrichment (Izzo et al., 2015; Tajitsu et al., 2015; Izzo et al., 2019; Molaro et al., 2020b). Also, the spectroscopic surveys have resulted in the detection and better characterization of large samples of Li-rich and super-Li-rich giants and highlighted their importance as Li producers (see, e.g., Martell et al., 2020; de la Reza, 2020; Deepak and Reddy, 2020; Kumar et al., 2020; and references therein).

At the same time, thanks to the numerous spectroscopic observations, the empirical evolution of Li with metallicity has been much better constrained, also separating the distribution in the different Galactic components. As far as the thin disc is concerned, many recent studies based on field star samples in the solar vicinity have suggested that the upper envelope of the Li vs. [Fe/H] distribution, which should be representative of the original interstellar medium abundance, declines at supersolar metallicities (Delgado Mena et al., 2015; Guiglion et al., 2016;

Bensby and Lind, 2018; Fu et al., 2018; Stonkutė et al., 2020). On the theoretical side, it is difficult to explain and model this result (Grisoni et al., 2019), but it has been speculated that it may be due to reduced production in the metal-rich regime (Prantzos et al., 2017; Fu et al., 2018; Grisoni et al., 2019), for example, because of lower AGB yields and/or a lower occurrence of nova or binary systems at high metallicity. However, different studies (Cummings et al., 2012; Bensby and Lind 2018; Guiglion et al., 2019), instead, suggested that those findings might be affected by selection biases; in particular, metal-rich field stars in the solar vicinity would be old stars migrated from the inner parts of the disc, depleting lithium as they traveled and got older. In other words, Li in those stars may not be representative of the original ISM value.

Those latter suggestions were tested in a very recent paper by Randich et al. (2020). They proposed the idea that Li measurements in metal-rich, young populations, that have presumably not depleted any Li and hence retain their original Li content, would allow tracing the evolution of Li in the ISM at high metallicity; to this aim, they exploited GES observations of open clusters and, specifically, of cluster members with supposedly pristine unprocessed Li, representative of the ISM value (very young PMS stars or stars on the warm side of the dip). Their results are reported in **Figure 10**; specifically, they very clearly demonstrated that previous claims of a Li decline at supersolar metallicities were due to selection biases in the used samples and that Li abundance in the ISM instead does not decrease at high metallicity but may actually increase. The data presented by Randich et al. (2020) also suggested—for the first time—the presence of a mild trend of A(Li) with Galactocentric distance, or a shallow gradient, to be confirmed with the analysis of the full data set.

Interestingly, in the last few years, Li observations in the Galactic thick disc and even Bulge have been reported (Delgado Mena et al., 2015; Bensby and Lind 2018; Fu et al., 2018; Bensby et al., 2020; Stonkutė et al., 2020). The different studies on the thick disc are not in agreement: indeed, some studies claim that Li rises with metallicity; others find that the distribution is flat or even decreases. Bensby and Lind (2018) convincingly showed that these discrepancies are very likely due to how the thick disc samples were selected; in particular, using chemical abundance criteria to define the thin and thick disc samples may lead to contamination of thin-disk stars into the thick disk samples. They hence concluded that there is a steady decrease of Li with metallicity in the thick disc, in turn implying that it has not undergone significant Li production during the first few billion years in the history of the Galactic disc. This suggests that the sources of Li production have longer evolutionary timescales.

Gravitational microlensing events offer an excellent opportunity to obtain spectra of unevolved stars in the Galactic Bulge. The analysis of 91 such targets (dwarfs and subgiants) by Bensby et al. (2020) has shown that old (age > 8 Gyr) Bulge stars with subsolar metallicities behave similarly to the thick disc and do not show any sign of Li production.

As for beryllium and boron, little progress has been made in the last few years. We just cite Molaro et al. (2020a) who collected lithium and beryllium measurements from the literature to investigate their trends in the Gaia-Enceladus Galaxy (Helmi et al., 2018). They found that the Be behavior for low-metallicity stars is similar to the Galactic one, while at somewhat higher metallicity ([Fe/H] > −1), the slope is shallower and the relationship is characterized by a smaller dispersion. On the one hand, this confirms that the scatter seen among MW stars is likely due to the fact that different components had been considered, which are mixed up in the Galactic halo; on the other hand, this may mean a reduced production of Be (or enhanced production of Fe) in Gaia-Enceladus (see discussion in Molaro et al., 2020a). In any case, this pilot result clearly highlights the usefulness of Be, along with *Gaia* data, to investigate in detail the formation of the Galactic halo and discs.

### 4.4.2 The Galactic Evolution of CNO

The abundance ratio trends of C, N, and O over Fe and over H vs. metallicity have profound implications for our understanding of the evolutionary status of the Galactic systems in which these elements are observed, since these elements are produced on different time scales and by stars with different masses (e.g., Chiappini et al., 2003, Chiappini et al., 2005; Cescutti et al., 2009; Vincenzo and Kobayashi, 2018a; Limongi and Chieffi, 2018; Prantzos, 2019). Galactic evolution of C, N, and O can be studied by determining their abundances in main-sequence F-G-K stars, spanning large ranges of ages and metallicities. Their atmospheres still essentially preserve the initial chemical composition of their parent molecular cloud, and thus, combining stars of different ages, they can trace the Galactic chemical evolution of these elements. For oxygen, we can also use abundances in evolved giant stars, since the surface abundance of this element is only slightly modified by stellar evolution, and photospheric abundances of evolved stars maintain the original oxygen abundance. Other good traces of the spatial distribution of CNO abundances are Cepheid stars, with their well-constrained distances (see, e.g., Lemasle et al., 2013; Genovali et al., 2015; Maciel and Andrievsky, 2019).

An alternative way to stellar spectroscopy is to measure the chemical composition of H II regions and planetary nebulae from their optical and ultraviolet emission-line spectra, both collisional excited lines, for oxygen and nitrogen, and recombination lines, for carbon and oxygen (e.g., Deharveng et al., 2000; Esteban et al., 2017, Esteban et al., 2018, Esteban et al., 2019; Stanghellini and Haywood, 2018; Arellano-Córdova et al., 2020; Esteban et al., 2020; Stanghellini et al., 2020). Abundances of H II regions and planetary nebulae are excellent probes of the radial abundance gradient of the Milky Way (e.g., Esteban et al., 2017, Esteban et al., 2018; Arellano-Córdova et al., 2020; Stanghellini et al., 2020). For a discussion on the methods and on the challenging issues on nebular abundances, we refer to Kewley et al. (2019) and García-Rojas (2020). In addition, the recent review of Maiolino and Mannucci (2019) summarizes the use of N/O and C/O abundance ratios as tracers of the evolution of the Milky Way and of nearby galaxies.

In this review, we focus on the recent results based on stellar spectroscopy in our Galaxy, for which we present some recent works making use of N/O and C/O abundance ratios to infer insights on the formation of our Galaxy. The wide range of ages spanned by stars in the Milky Way field populations and in clusters allows us, indeed, to trace the chemical evolution of these elements.

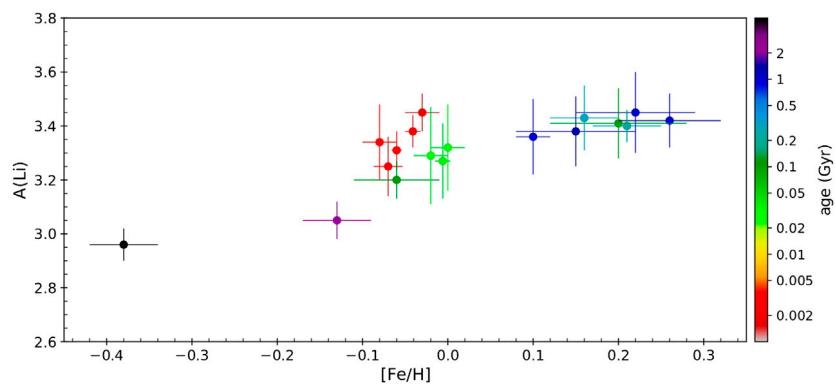

**FIGURE 10** | Average maximum Li abundance for the open clusters considered in the sample of Randich et al. (2020) as a function of the cluster metallicity. Clusters are color-coded by age [figure reproduced from Randich et al. (2020)].

In this framework, the abundance ratio N/O is a useful tool to study the interplay between Galactic processes (e.g., star formation efficiency, infall timescale, and outflow loading factor), thanks to the different origins and time scales of N and O (e.g., Vincenzo et al., 2016; Vincenzo and Kobayashi, 2018a; Magrini et al., 2018). In particular, the production of N in LIMS stars makes N/O a sort of clock to measure the time interval from the most recent star formation burst (e.g., Henry et al., 2000; Mollá et al., 2006). In addition, the evolution of the N/O ratio in galaxies, from low to high metallicity, is essential to understand its mechanisms of production and to validate some methods of measurement of O/H, which implicitly depend on the N/O ratio (cf. Berg et al., 2019; Roy et al., 2020). In a recent work, Magrini et al. (2018) discussed the abundances of N and O in open clusters and in field stars belonging in different portions of the thin and thick discs, showing a wide range of N/O, with values typical of the secondary production of N. This behavior is comparable to the one of H II regions located at different Galactocentric distances in external massive galaxies (see, e.g., Bresolin et al., 2004). There is also a general agreement with the global behavior of N/O vs. metallicity shown by a large sample of galaxies in the Local Universe (Belfiore et al., 2017). It is shown in **Figure 11**, where in the inner panel, we plot log (N/O) vs. [O/H] as a function of both stellar mass and radius in the sample of resolved galaxies (Belfiore et al., 2017), and in the main panel, we show log (N/O) vs. [Fe/H] in the Milky Way samples of field stars and open clusters of Magrini et al. (2018). Comparing the observed N/O vs. [Fe/H] in both cluster and field stars, with a grid of chemical evolution models, based on the models of Vincenzo et al. (2016), has provided further constraints to the inside-out formation of the Galactic disc, confirming a radial variation of the star formation and of the infall and outflow rates.

Also, the C/O ratio can be used to investigate the scenarios of the formation of our Galaxy. Recently, Amarsi et al. (2019) performed new calculations of C and O abundances in a sample of Galactic stars, considering NLTE and 3D effects, which improved the quality of the abundances and reduced the scatter. They found higher values of [C/O] at a given [O/H] in thin-disk stars with respect to the thick disc ones. The observed separation in the composition of the thin disc and of the thick disc is likely due to two main infall episodes, which happened in the Milky Way (Chiappini et al., 1997). In addition, thanks to their high-quality abundances, they detect an underdensity in the [C/O] vs. [O/H] plane, which might correspond to the onset of the second infall episode (see, e.g., Romano et al., 2019).

## 5 NEW PERSPECTIVES

Most of the forthcoming surveys and multiobject spectrographs, like 4MOST (de Jong et al., 2012), MOONS (Cirasuolo et al., 2011; Cirasuolo and Consortium 2020; Gonzalez et al., 2020), and WEAVE (Dalton et al., 2012), will focus on the optical and near-infrared range. MOONS will allow the measurement of chemical abundances, including CNO elements (Cirasuolo et al., 2014), in its high-resolution mode, observing two spectral regions at R∼20,000 (within the J- and H-bands). The 4MOST instrument will carry out several surveys focusing on stellar objects, aiming at performing Galactic archaeology of different components of the Milky Way and the Magellanic Clouds (see, e.g., Feltzing et al., 2018). The high-resolution mode R∼20,000 enables accurate abundance measurements of about 15 elements. In the spectral ranges of 4MOST, the G-band of CH (429–432 nm) will be included from which it is possible to derive the abundance of carbon, the CN band at 414–422 nm for the N abundance, and the atomic line [OI] at 630 nm for oxygen. The Li doublet at 670.8 nm will also crucially be covered. Similar bands and atomic lines will be observed in the high-resolution mode of WEAVE (R∼20,000), with two spectral windows in blue/green and in red (blue 404–465 nm or green 473–545 nm arms and red arm 595–68 nm).

Complementary to this, new planned instruments will enrich our knowledge of the light-element abundances in different ways. For example, MAVIS, an adaptive-optics assisted imager and spectrograph designed for VLT (McDermid 2019), will have a high-resolution mode both in blue and in red (R∼10,000–12,000), enabling the measurement of C and N from molecular bands and possibly Li in resolved stellar populations beyond the Milky Way.

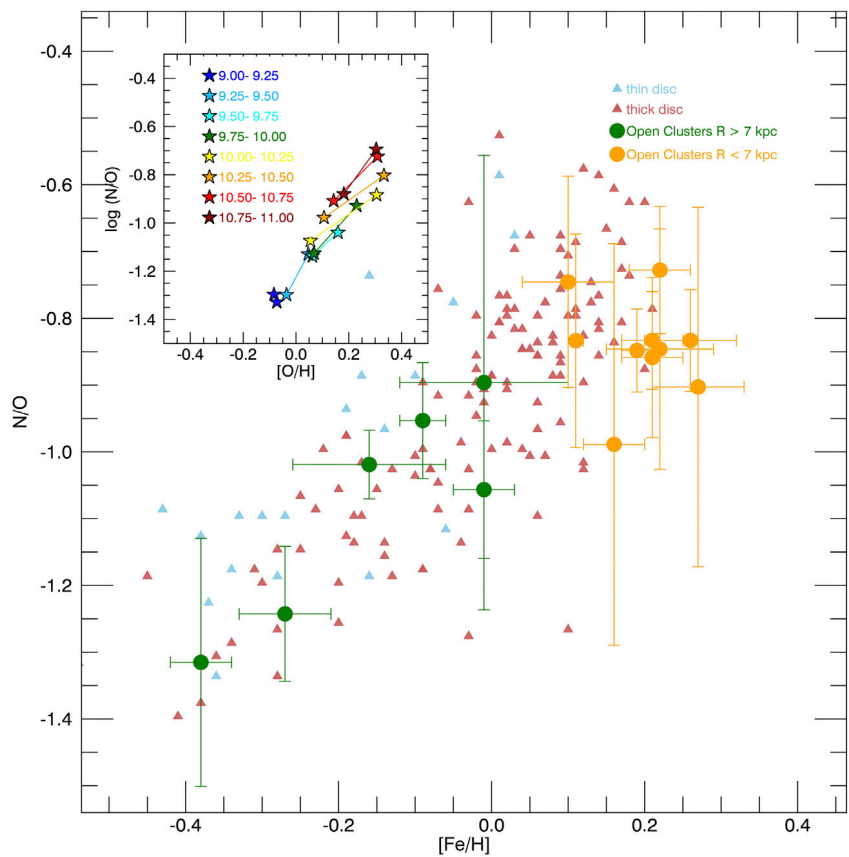

**FIGURE 11 |** log (N/O) vs. [Fe/H] in the Milky Way samples of field stars (blue and pink triangles and thin and thick discs, respectively) and of open clusters (in green clusters with $R_{GC} \geq 7$ kpc and in orange clusters with $R_{GC} \leq 7$ kpc) of Magrini et al. (2018). Inner panel: log (N/O) vs. [O/H] as a function of both stellar mass and radius in the sample of resolved galaxies of Belfiore et al. (2017). For each mass bin (see the legend for the range in the logarithm of the stellar mass of each bin), the uppermost star represents the innermost radial bin while the lower star represents the outermost radial bin.

Crucially, near-UV spectroscopy on the VLT will also be possible in the next few years with CUBES (Barbuy et al., 2014; Smiljanic, 2020). This will be a high-throughput, medium resolution (R ~ 20,000) spectrograph, operating between 300 and 400 nm. Phase A of the instrument started in Summer 2020. The instrument will allow us to observe significantly fainter objects than previously possible in the near-UV, opening up a new parameter space and allowing Be measurements in a variety of populations, including open and globular clusters. At the same time, CUBES measurements of CNO abundances from CN, NH, and OH bands will be possible for the same objects, offering excellent opportunities to further investigate the evolution of those elements (and in particular Be vs. O) based on homogeneous abundance determinations.

Finally, we mention that a concept study for a very high-resolution multiobject spectrograph to be put on the ESO VLT has been started. Such an instrument may indeed allow measurements of key isotopes in cluster and MW field stars.

## AUTHOR CONTRIBUTIONS

SR is responsible for the sections and discussion dedicated to Li, Be, and B, while LM focused on the review of C, N, and O. Both authors equally participated in the elaboration of the manuscript.

## FUNDING

LM and SR acknowledge the funding from MIUR Premiale 2016: Mitic. LM acknowledges the funding from the INAF PRIN-SKA 2017.

## ACKNOWLEDGMENTS

The data used in the review have been obtained from the GES Survey Data Archive and prepared and hosted by the Wide Field Astronomy Unit, Institute for Astronomy, University of Edinburgh, which is funded by the UK Science and Technology Facilities Council (STFC).

**Conflict of Interest:** The authors declare that the research was conducted in the absence of any commercial or financial relationships that could be construed as a potential conflict of interest.